\documentclass[%
 reprint,
 amsmath,amssymb,
 aps,
]{revtex4-2}

\usepackage{graphicx}
\usepackage{dcolumn}
\usepackage{bm}

\usepackage{xcolor}

\begin{document}

\preprint{APS/123-QED}

\title{Investigation of Ga interstitial and vacancy diffusion in $\beta$-Ga$_2$O$_3$ via split defects: \\ a direct approach via master diffusion equations}

\author{Channyung Lee}
    \affiliation
    {Department of Mechanical Science and Engineering, University of Illinois at Urbana-Champaign, 1206 W. Green Street, Urbana, Illinois 61801, United States.}
\author{Michael A. Scarpulla}
    \affiliation
    {Department of Materials Science and Engineering, University of Utah, Salt Lake City, Utah 84112, United States}
\author{Elif Ertekin}
    \affiliation
    {Department of Mechanical Science and Engineering, University of Illinois at Urbana-Champaign, 1206 W. Green Street, Urbana, Illinois 61801, United States.}
    \affiliation
    {Materials Research Laboratory, University of Illinois at Urbana-Champaign, Urbana, Illinois 61801}
    \email{ertekin@illinois.edu}

\date{\today}

\begin{abstract}

The low symmetry of monoclinic $\beta$-Ga$_2$O$_3$ leads to elaborate intrinsic defects, such as Ga vacancies split amongst multiple lattice sites. 
These defects contribute to fast, anisotropic Ga diffusion, yet their complexity makes it challenging to understand dominant diffusion mechanisms. 
Here, we predict the 3D diffusivity tensors for Ga interstitials (Ga${_\text{i}^{3+}}$) and vacancies (V${_{\text{Ga}}^{3-}}$) via first principles and direct solution of the master diffusion equations. 
We first explore the maximum extent of configurationally complex ``$N$-split'' Ga interstitials and vacancies.  
With dominant low-energy defects identified, we enumerate all possible elementary hops connecting defect configurations to each other, including interstitialcy hops. 
Hopping barriers are obtained from nudged elastic band simulations. 
Finally, the comprehensive sets of (i) defect configurations and their energies and (ii) the hopping barriers that connect them are used to construct the master diffusion equations for both
Ga${_\text{i}^{3+}}$ and V${_{\text{Ga}}^{3-}}$. 
The solution to these equations yields the Onsager transport coefficients, i.e.\ the components of the 3D diffusivity tensors 
$D_{\text{Ga}_{\text{i}}}$ and $D_{\text{V}_{\text{Ga}}}$ for Ga${_\text{i}^{3+}}$ and V${_{\text{Ga}}^{3-}}$, respectively. 
It further reveals the active diffusion paths along all crystallographic directions. 
We find that both Ga${_\text{i}^{3+}}$ and V${_{\text{Ga}}^{3-}}$ diffusion are fastest along the $c$-axis, due to 3-split defects that bridge neighboring unit cells along the $c$-axis and divert diffusing species around high-energy bottlenecks. 
Although isolated Ga${_\text{i}^{3+}}$ diffuse faster than isolated  V${_{\text{Ga}}^{3-}}$, self-diffusion of Ga is predominantly mediated by V$_{\text{Ga}}^{3-}$ due to the higher V$_{\text{Ga}}^{3-}$ defect concentration under most  thermodynamic environments.

\end{abstract}

\maketitle

\section{\label{sec:level1}Introduction}

The growing demand for power electronics requires the exploration of new semiconductor materials to effectively meet increasing needs for performance. 
Monoclinic gallium oxide ($\beta$-Ga$_2$O$_3$) is a notable candidate due to its unique properties, including an ultra-wide band gap ($\sim$4.8 eV), high breakdown voltage ($\sim$8 MV/cm), and tunable n-type conductivity \cite{1_Varley_2022,2_Fujita_2015,3_McCluskey_2020,0_Lee_2023}. 
Additionally, the availability of $\beta$-Ga$_2$O$_3$ in high-quality single crystal wafers and epitaxial thin films enhances its appeal for device fabrication and characterization \cite{4_Higashiwaki_2017,5_Higashiwaki_2016,6_Kim_2017}.

An understanding of the migration mechanisms of intrinsic defects is essential to exploit the potential of $\beta$-Ga$_2$O$_3$ in practical applications. 
The self-diffusion of Ga cations, for example, is mediated by Ga interstitials (Ga${_\text{i}}$) and vacancies (V${_{\text{Ga}}}$), and is a fundamental mechanism for mass transport within $\beta$-Ga$_2$O$_3$.
Ga${_\text{i}}$ and V${_{\text{Ga}}}$ native defects also can serve as vehicles for dopant diffusion, including shallow donors such as Si$_{\text{Ga}}$ and Sn$_{\text{Ga}}$, as well as deep compensating acceptors like Mg$_{\text{Ga}}$ and Fe$_{\text{Ga}}$ \cite{7_Azarov_2021,8_Mauze_2021,9_Peelaers_2019,10_Frodason_2023,11_Fahey_1989}.
Investigating migration mechanisms of Ga${_\text{i}}$ and V${_{\text{Ga}}}$ could lead to precise control over dopant distributions, enabling the tailoring of properties to improve performance and stability. 
Additionally, investigating transport mechanisms also helps elucidate degradation pathways like electromigration. 

The low symmetry of monoclinic $\beta$-Ga$_2$O$_3$ leads to many interesting complexes of intrinsic defects, such as Ga vacancies split between two or three neighboring Ga sites.
These 2, 3, and ``N''-split defects (N$= 2, 3, 4, ...$), many of which have been observed experimentally, are expected to contribute to fast and anisotropic ion diffusion in $\beta$-Ga$_2$O$_3$, posing challenges for understanding dominant diffusion mechanisms \cite{12_Johnson_2019,13_Kyrtsos_2017,14_Blanco_2005,15_Varley_2011,16_Frodason_2023,17_Ingebrigtsen_2018,18_Zimmermann_2020}. 
Previous computational studies, including a recent extensive analysis by Frodason \textit{et al.}\ \cite{10_Frodason_2023}, have explored migration pathways involving split defects. 
Yet, there remains an opportunity to fully elucidate the 3D diffusion network for both Ga${_\text{i}}$ and V${_{\text{Ga}}}$.  
A quantitative analysis of the diffusivity tensors $D_{\text{Ga}_{\text{i}}}$ and $D_{\text{V}_{\text{Ga}}}$, accounting for contributions from the full spectrum of intrinsic defects, would provide microscopic insights into defect migration pathways, diffusion anisotropy, and related processes like material degradation. 
Knowledge of the full diffusivity tensor obtained from first-principles may lead to predictions amenable to experimental validation.

In this study, we use first-principles calculations and a direct approach based on the solution of the master diffusion equations to determine the 3D diffusion tensors for Ga${_\text{i}^{3+}}$ and   V${_{\text{Ga}}^{3-}}$, interstitials and vacancies in their dominant charge state in $\beta$-Ga$_2$O$_3$. 
We first explore a wide range of defects, including $N$-split defects, and analyze (i) their formation energies and, (ii) the migration barriers for hops connecting one defect to another.
The analysis includes the identification of 32 unique interstitial and interstialcy hops between 20 different configurations of Ga interstitials, and 31 unique vacancy hops between 19 different configurations of Ga vacancies. 
By combining defect configuration energies and hopping barriers, we construct the master diffusion equations for both Ga${_\text{i}^{3+}}$ and  V${_{\text{Ga}}^{3-}}$ \cite{20_Trinkle_2016,21_Trinkle_2017,22_Trinkle_Onsager}. 
Solving these equations yields the Onsager transport coefficients in the form of 3D diffusivity tensors $D_{\text{Ga}_{\text{i}}}$ and $D_{\text{V}_{\text{Ga}}}$. 
We find that both Ga interstitials and Ga vacancies exhibit the highest diffusivity along the $c$-axis, and that the components of the interstitial diffusivity tensor $D_{\text{Ga}_{\text{i}}}$ are larger than those of vacancy diffusivity tensor $D_{\text{V}_{\text{Ga}}}$.
However, Ga self-diffusion is predicted to still be mediated by V${_{\text{Ga}}^{3-}}$ rather than Ga${_\text{i}^{3+}}$, due to the higher concentration of vacancies under typical environments. 
These findings are relevant to the design and optimization of $\beta$-Ga$_2$O$_3$ electronics.

\begin{figure*}[htbp!]
\centering
\includegraphics[width=7in]{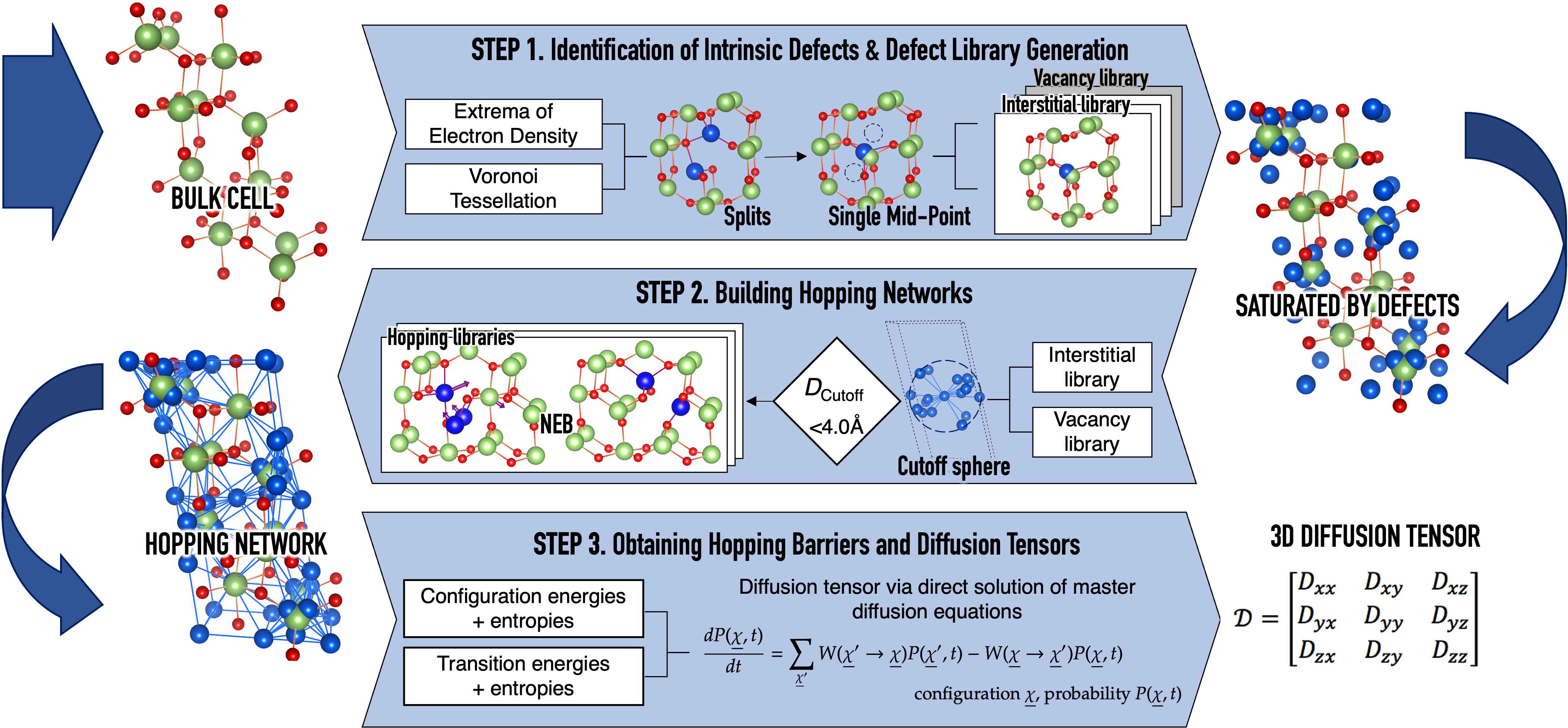}
\caption{\label{figure} 
Schematic representation of the workflow used in this work to obtain the diffusivity tensors for vacancy and interstitial defects in $\beta$-Ga$_2$O$_3$. The approach is based on assembling and solving the master diffusion equations via the Onsager methodology \cite{20_Trinkle_2016,21_Trinkle_2017,22_Trinkle_Onsager}. 
}
\end{figure*}

\section{Methods} 

\subsection{Master diffusion equations}

For assembling and then solving the master diffusion equations, we used the Onsager software package \cite{22_Trinkle_Onsager,20_Trinkle_2016,21_Trinkle_2017}, which yields the components of the diffusivity tensor. 
We implement the method separately for interstitial diffusion and vacancy diffusion, and obtain isolated diffusion tensors $D_{\text{Ga}_{\text{i}}}$ and $D_{\text{V}_{\text{Ga}}}$ respectively, for each mechanism.
Hence, we construct two independent diffusion networks, one for interstitials and the other for vacancies. 
The overall methodology consists of three stages, as illustrated in Figure 1. 

In the first stage (Figure 1, Step 1), defect libraries are generated by identifying possible atomic configurations of Ga${_\text{i}^{3+}}$ and  V${_{\text{Ga}}^{3-}}$, and determining their formation energies (site energies) as described below.
The defect library includes extended defect complexes such as $N$-split Ga${_\text{i}^{3+}}$ formed by combinations of $N$ interstitials and ($N$-1) vacancies, and $N$-split V${_{\text{Ga}}^{3-}}$ formed by combinations of $N$ vacancies and ($N$-1) interstitials.
For instance, a 4-split Ga${_\text{i}^{3+}}$ comprises four distinct Ga interstitials and three Ga vacancies, while a 4-split V${_{\text{Ga}}^{3-}}$ consists of four Ga vacancies and three Ga interstitials.
$N$-split Ga${_\text{i}^{3+}}$ and $N$-split V${_{\text{Ga}}^{3-}}$, respectively introduce an extra Ga or a missing Ga, resulting in a defect that is spread across multiple lattice sites instead of at a single lattice point (details in Figure S1). 
In the defect library, we employ a single representative mid-point located in the middle of a given $N$-split defect to represent the position of each $N$-split defect, facilitating the tracking of transitions between extensive $N$-split configurations and accommodating interstitialcy hopping.

In the second stage (Figure 1, Step 2), the complete hopping network is assembled.  
All symmetry-unique interstitial-to-interstitial hops (for the interstitial diffusion network) and vacancy-to-vacancy hops (for the vacancy network), lying within a 4 {\AA} cutoff distance, are enumerated.
We then obtain migration barriers for these hops using the climbing image-nudged elastic band (CI-NEB, see below). 
The library of defect configurations (and associated site energies) and the library of hops (and associated transition rates) that connect different configurations to each other form the three-dimensional diffusion network for each mechanism. 

In the final stage (Figure 1, Step 3), the identified diffusion pathways, site energies, and transition rates are assembled into a system of coupled rate equations. 
The formulation is based on the Onsager reciprocal relations that express the linear relationship between generalized fluxes and forces in thermodynamic systems. The constants of proportionality (the Onsager coefficients) are the diffusivities. 
The Onsager formulation relies on the assumption of well-defined configurations $\chi$ that thermalize faster than transitions occur. 
Diffusion is considered as a Markov process consisting of transitions from one state $\chi$ to another $\chi'$. 
The quantity $P(\chi,t)$ denotes the probability of finding the system in state $\chi$ at time $t$. 
As indicated in Figure 1, Step 3, the time rate of change is given by  
{\small
\begin{equation}
\frac{dP(\chi,t)}{dt} = \sum_{\chi'}\left(W(\chi' \rightarrow \chi) P(\chi',t)-W(\chi \rightarrow \chi') P(\chi,t)\right) \hspace{0.5em},  
\end{equation}
}
involving a sum over transition rates $W(\chi' \rightarrow \chi)$ from all other configurations $\chi'$ into configuration $\chi$ and a sum over transition rates $W(\chi \rightarrow \chi')$ from configuration $\chi$ into all other configurations $\chi'$. 
Under thermodynamic equilibrium where $P(\chi,t)$ is fixed, the system of equations above (one for each configuration $\chi$) can be solved under the constraint of detailed balance. 
By assembling and self-consistently solving the system, we obtain the linear transport coefficients \(D_{\text{Ga}_{\text{i}}^{3+}}\) and \(D_{\text{V}_{\text{Ga}}^{3-}}\), the diffusivity tensors for isolated $\text{Ga}_{\text{i}}^{3+}$ and $\text{V}_{\text{Ga}}^{3-}$, respectively.
Finally, self-diffusion coefficients for Ga, labeled \(D_{\text{Self}, \text{Ga}_{\text{i}}^{3+}}\) and \(D_{\text{Self}  \text{V}_{\text{Ga}}^{3-}}\) respectively for interstitials and vacancies, are obtained using jump balance.
A more detailed description of the approach is given in Refs.\ \cite{22_Trinkle_Onsager,20_Trinkle_2016,21_Trinkle_2017}; for completeness a brief discussion is presented in Appendix A. 

\subsection{Site energies and migration barriers}

To obtain site energies and migration barriers, we used first-principles simulations. 
All first-principles simulations were performed using density functional theory \cite{23_Hohenberg_1964,24_Kohn_1965} with the projector augmented wave (PAW) method \cite{25_Bloch_1994,26_Kresse_1999} as implemented in the Vienna \textit{Ab Initio} Simulation Package (VASP) \cite{27_Kresse_1996,28_Kresse_1996_2}.  
The Perdew–Burke–Ernzerhof (PBE)\cite{29_PBE_1996} parametrization of the generalized gradient approximation (GGA) \cite{30_GGA_1992} was used to describe the exchange-correlation functional.
The plane-wave basis cutoff was set at 420 eV, and Ga 3d electrons were explicitly included as valence states in the chosen pseudopotentials.
For geometry optimization, the convergence criteria were set at $1\times10^{-6}$ eV for energy and 0.001 eV/{\AA} for the residual forces on each atom.
The ground-state lattice parameters of the monoclinic $\beta$-Ga$_2$O$_3$ conventional unit cell were determined to be $a =$ 12.47 {\AA, $b =$ 3.09 {\AA}, $c =$ 5.88 {\AA}, and $\beta =$ 103.7$^{\circ}$
These values agree well with previously reported results obtained using PBE functionals \cite{13_Kyrtsos_2017,40_Yoshioka_2007,33_Zacherle_2013} and experimental measurements \cite{34_Geller_1960,35_Ahman_1996}.

The focus of this work is on the 3+ charge state of Ga$_{\text{i}}$ and the 3- charge state of V$_{\text{Ga}}$, the stable charge state for each defect under typical n-type doping or unintentionally doped conditions in $\beta$-Ga$_2$O$_3$.
Defect formation energies (site energies) were obtained using the usual supercell formulation, as described in Appendix B.
To accurately describe the extended $N$-split defects when determining their formation energies, we employed different supercell sizes at different stages of the study. 
In the initial stage, we utilized 160 atoms in 1$\times$4$\times$2 supercells to comprehensively screen and identify all potential defective configurations which we assembled from multiple sources. 
The configurations analyzed include point interstitials obtained through Voronoi tessellations \cite{36_Goyal_2017} and electron density topology analysis \cite{14_Blanco_2005}, as well as new $N$-split defects designed based on generalization of previously reported 2 or 3-split defects \cite{16_Frodason_2023,13_Kyrtsos_2017,15_Varley_2011}. 
In total, we considered 27 unique Ga${_\text{i}}$ structures and 25 unique V${_{\text{Ga}}}$ structures. 
After excluding the highest energy structures, we used 1$\times$4$\times$3 supercells with 240 atoms and 2$\times$4$\times$2 supercells with 320 atoms to minimize finite size effects from the extended nature of $N$-split defects along the $c$-axis and $a$-axis, respectively. 
A 2$\times$2$\times$2 k-point grid generated by the Monkhorst-Pack method was used for all supercells \cite{31_Monkhorst_1976}.

Migration barriers were calculated using the climbing-image nudged elastic band \cite{19_Henkelman_2000} with a 0.1 eV/{\AA} convergence criterion for the residual forces on each atom.
Considering the computational cost of NEB calculations, for a given defect we selected one from four different possible supercell sizes (1$\times$4$\times$2, 1$\times$4$\times$3, 2$\times$4$\times$2, and 2$\times$4$\times$3) based on the crystal directions along which starting and ending structures are most extended, again to minimize finite size effects.

To account for uncertainties in estimated migration barriers arising from the choice of PBE for the DFT exchange correlation functional, we benchmarked the current PBE migration barriers using 1$\times$4$\times$2 and 1$\times$4$\times$3 supercells against the results from Frodason et al.\ \cite{16_Frodason_2023}, who employed both the Strongly Constrained and Appropriately Normed (SCAN) \cite{53_SCAN} and Heyd-Scuseria-Ernzerhof (HSE) \cite{54_HSE_1,55_HSE_2} functionals.
This comparison revealed a mean discrepancy of 0.085 eV for the current PBE results and their HSE-calculated barriers. 
This discrepancy is similar to the differences between SCAN and HSE reported by Frodason et al.\, suggesting that PBE and SCAN give similar barriers for this material system. 
Therefore, in the forthcoming results, we show predicted diffusivities in a range of values accounting for a $\pm$0.1 eV uncertainty for all migration barriers.  

\begin{figure*}[htbp!]
\centering
\includegraphics[width=7in]{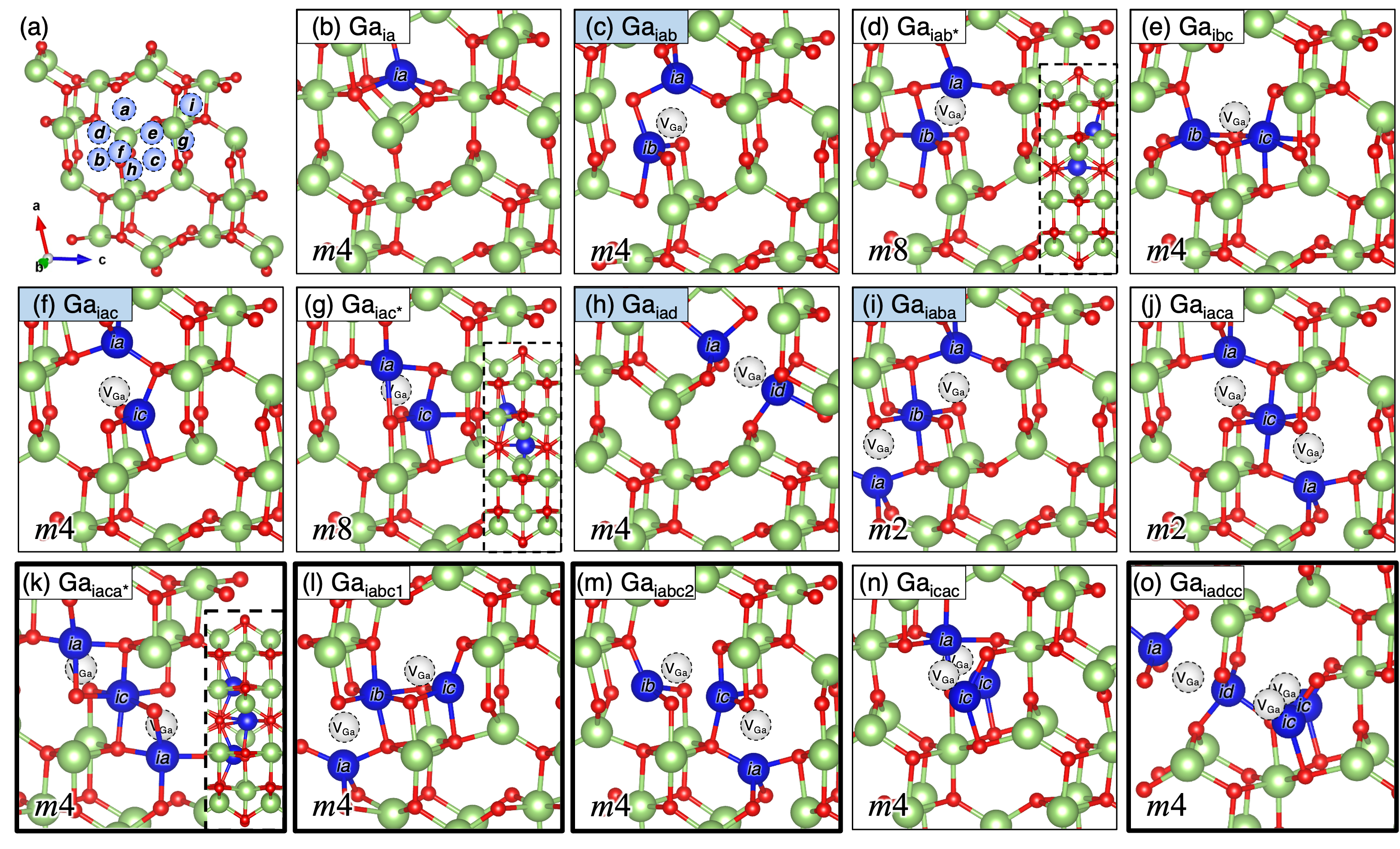}
\caption{
(a) Monoclinic $\beta$-Ga$_2$O$_3$ structure, highlighting unique Ga interstitial sites ($a$-$i$).
A series of relaxed defective $\beta$-Ga$_2$O$_3$ structures with selected Ga$_{\text{i}}^{3+}$ defects including (b) a single interstitial, (c-h) 2-split interstitials, (i-n) 3-split interstitials, and (o) a 4-split interstitial. 
The Greek letter in the lower left corner denotes the multiplicity, $m$, of Ga$_{\text{i}}^{3+}$ defects within the unit cell. 
The newly identified structures are indicated by a bold outline. 
Name boxes colored in blue are used to indicate the four structures with the lowest energy. The interstitials shown are the ones predicted to contribute to interstitial-mediated diffusion. 
}
\end{figure*}

\section{Results and Discussion}

\subsection{Defect configurations and formation energies}

Previous theoretical studies have revealed diverse ranges of energetically stable defect complexes in $\beta$-Ga$_2$O$_3$. 
Therefore our first objective was to systematically explore and identify as many as stable defect structures in $\beta$-Ga$_2$O$_3$ as possible, including extended defects.
We tested numerous combinations of split defect structures by further splitting along the $ac$ plane.
This approach was motivated by previous theoretical findings that the tetrahedral Ga$_{\text{(I)}}$ atoms may shift towards the center of the $b$-axis channels through the formation of stable octahedral or tetrahedral Ga-O bonds \cite{17_Ingebrigtsen_2018,16_Frodason_2023,15_Varley_2011}. 
This shift induces the formation of split defects such as Ga$_{\text{i}}$-V$_{\text{Ga}}$-Ga$_{\text{i}}$ 2-split interstitials (comprising two interstitials and one vacancy) or V$_{\text{Ga}}$-Ga$_{\text{i}}$-V$_{\text{Ga}}$ 2-split vacancies (comprising two vacancies and one interstitial), and possibly further extended defects on larger scales.
Specifically, the rearrangement of multiple Ga$_{\text{(I)}}$ atoms allows for the formation of continuous defect chains consisting of $N$ vacancies and ($N$-1) interstitials (or vice versa).
Through our exploration, we identified a total of 25 distinct Ga$_{\text{i}}^{3+}$ configurations, ranging from single-point interstitials to 4-split interstitials, and a total of 27 distinct V$_{\text{Ga}}^{3-}$ configurations, ranging from single-point vacancies to 4-split vacancies.

To facilitate structural characterization of a wide range of defect complexes, here onwards we employ a simplified notation ($a$-$i$) to label the unique possible lattice sites for Ga atoms in the $\beta$-Ga$_2$O$_3$ unit cell, as depicted in Figure 2(a). 
The sites are illustrated from the $b$-axis channel viewing direction, which we found to be the easiest direction to observe the configurations. 
The lattice site located in the large eight-sided channel (A channel) is denoted as $a$, while the sites positioned in the two irregular hexagonal channels (B and C channels) are labeled as $b$ and $c$, according to established conventions \cite{17_Ingebrigtsen_2018,10_Frodason_2023}. 
Additionally, the lattice sites located between the tetrahedral Ga$_{\text{(I)}}$ and the hexagonal Ga$_{\text{(II)}}$ are denoted as $d$ and $e$, while those located between two tetrahedral Ga$_{\text{(I)}}$ and between two hexagonal Ga$_{\text{(II)}}$ along the $b$-axis are labeled as $f$ and $g$, respectively.
These notations for lattice sites are used to describe both $N$-split interstitials and $N$-split vacancies.

\subsubsection{Ga Interstitial Configurations}

The main interstitial Ga$_{\text{i}}^{3+}$ configurations that we ultimately find to be responsible for the formation of the major diffusion pathways along each of the three different crystal orientations ($a^*$, $b$, and $c$) are shown in Figure 2(b-o); the additional configurations considered can be found in Figure S2. 
The newly identified structures are indicated by a bold outline, while the four structures with the lowest energies are highlighted by blue-colored name boxes. 
First, we investigated three distinct configurations of single interstitials positioned in the A, B, and C channels. 
The structure of Ga$_{{\text{ia}}}$ in the A channel, as shown in Figure 2(b), exhibits stable pyramidal Ga-O bonds. 
On the other hand, Ga$_{{\text{ib}}}$ and Ga$_{{\text{ic}}}$, located in the B and C channels, respectively (shown in Figure S2 (a,b)), form octahedral Ga-O bonds by attracting four neighboring threefold oxygen atoms (O${_{\text{ii}}}$ and O${_{\text{i}}}$, respectively) towards the central Ga interstitial, resulting in significantly higher energies compared to Ga$_{{\text{ia}}}$.

For N-split interstitials, we identified unique combinations of $N$ interstitials split into multiple sites ($a$-$g$) and the corresponding host ($N$-1) Ga vacancies that arise when simple interstitials shift.
Regarding 2-split interstitials, we identified a total of 8 distinct structures.
The 2-split interstitials Ga$_{{\text{iab}}}$, Ga$_{{\text{ibc}}}$, and Ga$_{{\text{iac}}}$ (Figure 2(c,e,f)) exhibit Ga$_{\text{i}}$-V${_{\text{Ga}}}$-Ga$_{\text{i}}$ structures, where two interstitials are positioned in three distinct channels sharing one V${_{\text{Ga(I)}}}$ site. 
We also tested another configuration, Ga$_{{\text{iah}}}$ (Figure S2(d)), in which the interstitial is split between site $a$, and, at a location halfway between B and C channels. 
However, this configuration was found to have an energy of more than 1 eV higher than the others. 
Additionally, we explored other sets of 2-split interstitials denoted as Ga$_{{\text{iad}}}$ and Ga$_{{\text{iei}}}$ (Figure 2(h) and Figure S2(c), respectively), which share the V${_{\text{Ga(II)}}}$ site and for which the two Ga$_i$ form tetrahedral bonds in different orientations. 
However, the presence of an interstitial atom in the small rhombohedral channel in Ga$_{{\text{iei}}}$ significantly distorts the lattice, resulting in high energy compared to Ga$_{{\text{iad}}}$.
Among the investigated 2-split structures, the Ga$_{{\text{iac}}}$ configuration exhibited the lowest formation energy, while Ga$_{{\text{iad}}}$ showed the second lowest. 
During NEB calculations of $b$-axis hops involving Ga$_{{\text{iab}}}$ and Ga$_{{\text{iac}}}$, we discovered two additional non-symmetric structures along the $b$-axis, denoted Ga$_{{\text{iab}^*}}$ and Ga$_{{\text{iac}^*}}$ (Figure 2(d,g)). 
In these structures, the interstitials Ga$_{{\text{ia}}}$ exhibit slight displacements from the $ac$ plane where the V$_{\text{Ga(I)}}$ is located, compared to the corresponding symmetric structures of Ga$_{{\text{iab}}}$ and Ga$_{{\text{iac}}}$ shown in Figure 2(d,g), respectively. 

We further identified 15 unique configurations of 3-split interstitials. 
These were constructed by associating pairs of two nearby Ga vacancies with three neighboring Ga interstitials: one situated between the two vacancies and the others adjacent to each vacancy.
From the V${_{\text{Ga(I)}}}$-V${_{\text{Ga(I)}}}$ pair centered on channel B of the $ac$ plane, we identified Ga$_{{\text{iaba}}}$, Ga$_{{\text{icbc}}}$, and Ga$_{{\text{iabc1}}}$ (Figure 2(i,l), and Figure S2(i), respectively).
In the V${_{\text{Ga(I)}}}$-V${_{\text{Ga(I)}}}$ set centered in the C channel in the $ac$ plane, we found Ga$_{{\text{iaca}}}$, Ga$_{{\text{ibcb}}}$, and Ga$_{{\text{iabc2}}}$ (Figure 2(j,m), and Figure S2(h), respectively). 
Although previous studies often categorize Ga$_{{\text{iaba}}}$ and Ga$_{{\text{iaca}}}$ as point interstitials \cite{16_Frodason_2023}, we classified these configurations as 3-split interstitials because the two shifted interstitial atoms from Ga(I) sites form stable tetrahedral bonds in different A channels, deviating from the original hexagonal channels.
Along the $b$-axis, the V${_{\text{Ga(I)}}}$-V${_{\text{Ga(I)}}}$ set yielded Ga$_{{\text{icac}}}$ and Ga$_{{\text{ibab}}}$ (Figure 2(n) and Figure S2(k), respectively).
In the V${_{\text{Ga(I)}}}$-V${_{\text{Ga(II)}}}$ set, we identified Ga$_{{\text{iadc}}}$ and Ga$_{{\text{iaha}}}$ (Figure S2(j,m)).
Ga$_{{\text{ieie}}}$ was identified in the V${_{\text{Ga(I)}}}$-V${_{\text{Ga(I)}}}$ set (Figure S2(l)).
We were able to identify additional off-symmetric structures along the $b$-axis for 3-split interstitials, specifically derived from Ga$_{{\text{iaba}}}$ and Ga$_{{\text{iaca}}}$.
These additional structures, labeled Ga$_{{\text{iaba}^{*}}}$, Ga$_{{\text{iaba}^{**}}}$, Ga$_{{\text{iaca}^{*}}}$, and Ga$_{{\text{iaba}^{**}}}$ (Figure S2(e,f), Figure 2(k), and Figure S2(g), respectively), exhibit shifted $ia$ split interstitials along the $b$ axis, deviating from the $ac$-plane where the $ia$ and $ib$ interstitials are located along with two V${_{\text{Ga(I)}}}$'s, respectively.
Lastly, we identifed only one 4-split interstitial, Ga$_{{\text{iadcc}}}$ (Figure 2(o)), which exhibits a structure similar to Ga$_{{\text{iadc}}}$ and Ga$_{{\text{icac}}}$.
Due to the increased size of the defect cluster and resulting complexities, we were unable to extensively explore further 4-split interstitials. 

\begin{figure*}[htbp!]
\centering
\includegraphics[width=7in]{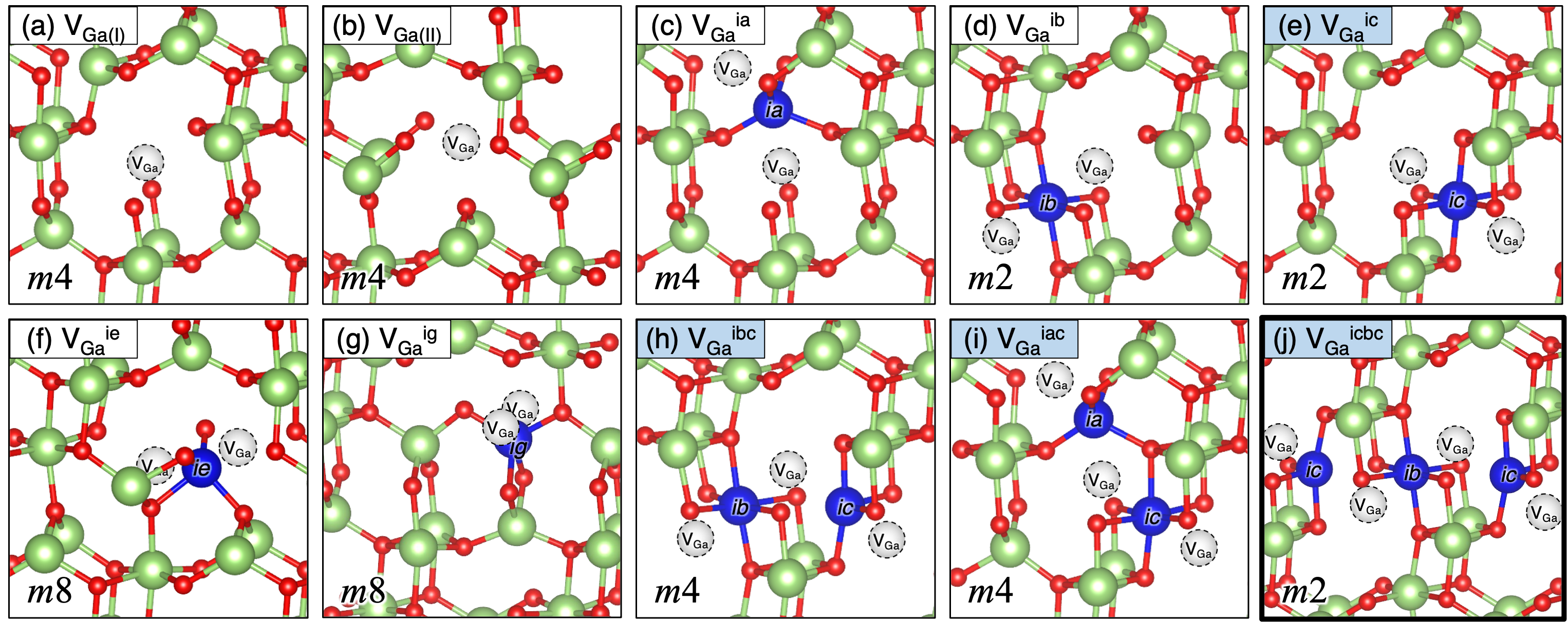}
\caption{
Relaxed defective $\beta$-Ga$_2$O$_3$ structures with selected V$_{\text{Ga}}^{3-}$ defects including (a, b) single point vacancies, (c-g) 2-split vacancies, (h, i) 3-split vacancies, and (j) a 4-split vacancy. The vacancies shown are the ones predicted to contribute to vacancy-mediated diffusion. 
}
\end{figure*}

After excluding high-energy structures based on a cutoff of 1 eV for relative formation energies, referenced to the lowest energy structure of Ga$_{{\text{iac}}}$ as shown in Figure S4(a), we increased the size of the supercells to accurately capture the extended nature of the $N$-split interstitials.
As a specific example, in the well-known Ga interstitial structures of Ga$_{{\text{iaba}}}$ and Ga$_{{\text{iaca}}}$, as shown in Figure 2(i,j), respectively, the split interstitial chains extend along the $a$-axis and interact directly with neighboring supercells above and below simultaneously.
Under 1$\times$4$\times$2 supercells, this interaction leads to spurious lattice distortions at the supercell boundary as two Ga(II) atoms shift toward the center of the lattice, resulting in high strain energy under periodic boundary conditions. 
Therefore, we reoptimized low-energy Ga$_{\text{i}}^{3+}$ structures and recalculated their formation energies using 1$\times$4$\times$3 and 2$\times$4$\times$2 supercells (Figure S4(b)). 

When expanding the supercells, for single interstitial Ga$_{{\text{ia}}}$, there were no noticeable differences in the formation energies. 
However, substantial formation energy variations were observed for $N$-split Ga interstitials depending on the supercell size, emphasizing the importance of selecting the appropriate supercell according to the direction of extension.
In 2$\times$4$\times$2 supercells, the formation energies of Ga$_{{\text{iab}}}$, Ga$_{{\text{iac}}}$, Ga$_{{\text{iad}}}$ decreased by approximately 0.1 eV each and those of Ga$_{{\text{iaba}}}$ and Ga$_{{\text{iaca}}}$ decreased by approximately 0.4 eV each when adequate spacing was included between defect clusters or chains.
In particular, significant energy differences were observed in Ga$_{{\text{iaba}}}$ and Ga$_{{\text{iaca}}}$, suggesting that their stability may have been underestimated due to finite size effects in previous theoretical calculations using conventional 1$\times$4$\times$2 supercells.
Similarly, using 1$\times$4$\times$3 supercells, the formation energies of Ga$_{{\text{ibc}}}$ and Ga$_{{\text{iadcc}}}$ decreased by approximately 0.15 eV each and those of Ga$_{{\text{ibcb}}}$ and Ga$_{{\text{icbc}}}$ decreased by approximately 0.4 eV each due to sufficient spacing between defect clusters along the $c$-axis.

\subsubsection{Ga Vacancy Configurations}

Figure 3 illustrates the relaxed V${_{\text{Ga}}^{3-}}$ structures, including point vacancies and $N$-split vacancies ultimately found to be important for vacancy diffusion. 
As before, newly identified structures are indicated by a bold outline, while the four structures with the lowest energies are highlighted by blue-colored name boxes. 
Consistent with previous studies, the formation energy of V${_{\text{Ga(I)}}}$ was found to be higher than V${_{\text{Ga(II)}}}$ (Figure 3(a,b)), indicating that V${_{\text{Ga(I)}}}$ serves as the more stable split vacancy center. 
To identify 2-split vacancies, we followed a similar approach used in the identification of $N$-split interstitials, examining V${_{\text{Ga}}}$-V${_{\text{Ga}}}$ sets and placing an interstitial between two vacancies.
A total of seven distinct 2-split vacancies were identified from each V${_{\text{Ga}}}$-V${_{\text{Ga}}}$ arrangement.
In the $ac$-plane, three different 2-split vacancies (V${_{\text{Ga}}^{\text{ia}}}$, V${_{\text{Ga}}^{\text{ib}}}$, and V${_{\text{Ga}}^{\text{ic}}}$) as illustrated in Figure 3(c,d,e) showed lower formation energies than the other off $ac$-plane 2-split vacancies, such as V${_{\text{Ga}}^{\text{ie}}}$, V${_{\text{Ga}}^{\text{ig}}}$, V${_{\text{Ga}}^{\text{id}}}$, and V${_{\text{Ga}}^{\text{if}}}$ (Figure 3(f,g) and Figure S4(a,b), respectively). 
To identify $N$-split vacancies where $N \geq 3$, we employed a more straightforward method based on the identified 2-split vacancies. 
This method involved coupling two adjacent low-energy 2-split vacancies (V$_{\text{Ga}}$Ga$_{\text{i}}$V$_{\text{Ga}}^*$ + V$_{\text{Ga}}^*$Ga$_{\text{i}}$V$_{\text{Ga}}$) in such a way that they share one vacancy site V$_{\text{Ga}}^*$, resulting in another split vacancy of V$_{\text{Ga}}$Ga$_{\text{i}}$V$_{\text{Ga}}^*$Ga$_{\text{i}}$V$_{\text{Ga}}$.
This strategy enabled us to create extended chains of (V$_{\text{Ga}}$Ga$_{\text{i}}$)$_n$V$_{\text{Ga}}$, representing continuous sequences of $N$-split vacancies.
As a result, we identified 7 and 9 unique triple and quadruple splits, respectively.

To mitigate finite-size effects, we again employed larger supercells, 1$\times$4$\times$3 and 2$\times$4$\times$2, to re-optimize the low-energy V${_{\text{Ga}}^{3-}}$ structures we identified.
As depicted in Figure S5, no significant differences are observed in the formation energies of the point vacancies and the 2-split vacancies.
This finding can be attributed to the small size of the 2-split vacancy clusters, which results from the localized lattice distortion caused by the vacancies and the confined interstitial within the clusters.

In contrast, we observed substantial variations in the formation energies of more extended 3-split and 4-split vacancies.
The relative formation energy of V${_{\text{Ga}}^{\text{ibc}}}$ (Figure 3(h)) decreased by 0.42 eV in 1$\times$4$\times$3 supercell, consistent with earlier findings by Frodason et al.\ \cite{16_Frodason_2023}. 
The formation energies of  V${_{\text{Ga}}^{\text{iab}}}$ and V${_{\text{Ga}}^{\text{iac}}}$ (Figure S3(c) and Figure 3(i), respectively) also decreased by approximately 0.17 and 0.11, respectively, with 2$\times$4$\times$2 supercells.
In the case of 4-split vacancies, a significant decrease in relative formation energies was observed for V${_{\text{Ga}}^{\text{iaba}}}$ and V${_{\text{Ga}}^{\text{iabc2}}}$ (Figure S3(i,l)) when using the 2$\times$4$\times$2 supercell (reduced by 0.37 eV and 0.09 eV, respectively), as well as for V${_{\text{Ga}}^{\text{ibcb}}}$, V${_{\text{Ga}}^{\text{iabc1}}}$, and V${_{\text{Ga}}^{\text{iabc3}}}$ (Figure S3(k,h,m)) when using the 1$\times$4$\times$3 supercell (reduced by 0.53, 0.39, and 0.36 eV respectively).

Among the identified vacancies, V${_{\text{Ga}}^{\text{ic}}}$ (Figure 2(e)) exhibited the lowest formation energy, followed by V${_{\text{Ga}}^{\text{iabc2}}}$ and V${_{\text{Ga}}^{\text{iaba}}}$ (Figure S2(i,l)) with relative energy differences of only 0.06 and 0.11 eV, respectively.
It should be noted that triple and 4-split vacancies showed even lower formation energies compared to well-known 2-split vacancies such as V${_{\text{Ga}}^{\text{ia}}}$ and V${_{\text{Ga}}^{\text{ib}}}$, as well as point vacancies V${_{\text{Ga(I)}}}$ and V${_{\text{Ga(II)}}}$, highlighting the possibility of significantly longer stable split vacancy chains extending across multiple unit cells. Such vacancy chains, in turn, would require even larger supercells to accurately characterize.

Recent studies have explored the possibility that $N$-split defects play a role in the phase transition between the $\beta$ and $\gamma$ phases of Ga$_2$O$_3$ \cite{42_Huang_2023,49_Huang_2023}.
The $\gamma$ phase, a metastable polymorph of Ga$_2$O$_3$ that exhibits the defect spinel structure, forms a disordered cation arrangement while sharing a similar anion lattice skeleton with the $\beta$ phase \cite{40_Yoshioka_2007,50_Charlotte_2024}.
Experimental investigations frequently observe the presence of $\gamma$-phase layers on the surfaces of $\beta$-Ga$_2$O$_3$ films grown under various conditions \cite{41_Chang_2021}.
The $\gamma$ phase structures resemble the split interstitials observed in the A, B, and C hexagonal channels of $\beta$-Ga$_2$O$_3$; these interstitials can arise as part of the $N$-split vacancies. 
In recent theoretical work, Huang et al.\ proposed the formation of Ga defect complexes that involve relaxations of multiple Ga defects, probably 2-split Ga vacancies, resulting in a local structure similar to the $\gamma$ phase \cite{42_Huang_2023}. 
This structural analogy suggests the possibility that larger-scale formation of $N$-split Ga vacancies, extending beyond 4-split throughout the bulk, could potentially induce a phase transition between the $\beta$ and $\gamma$ phase Ga$_2$O$_3$. 

\subsection{Diffusion networks and migration energy barriers}

In the following subsections, we first describe our approach and main findings for the interstitial and vacancy diffusion networks.
These findings include components of the vacancy and interstitial diffusion tensors, effective activation energies, discussion of anisotropy, estimates of Ga cation self-diffusion coefficients, and comparison to available experiments.
Finally, we subsequently break down the diffusion networks for interstitials and vacancies in detail to identify the dominant elementary hops and diffusion mechanisms. 

\begin{figure*}[htbp!]
\centering
\includegraphics[width=7in]{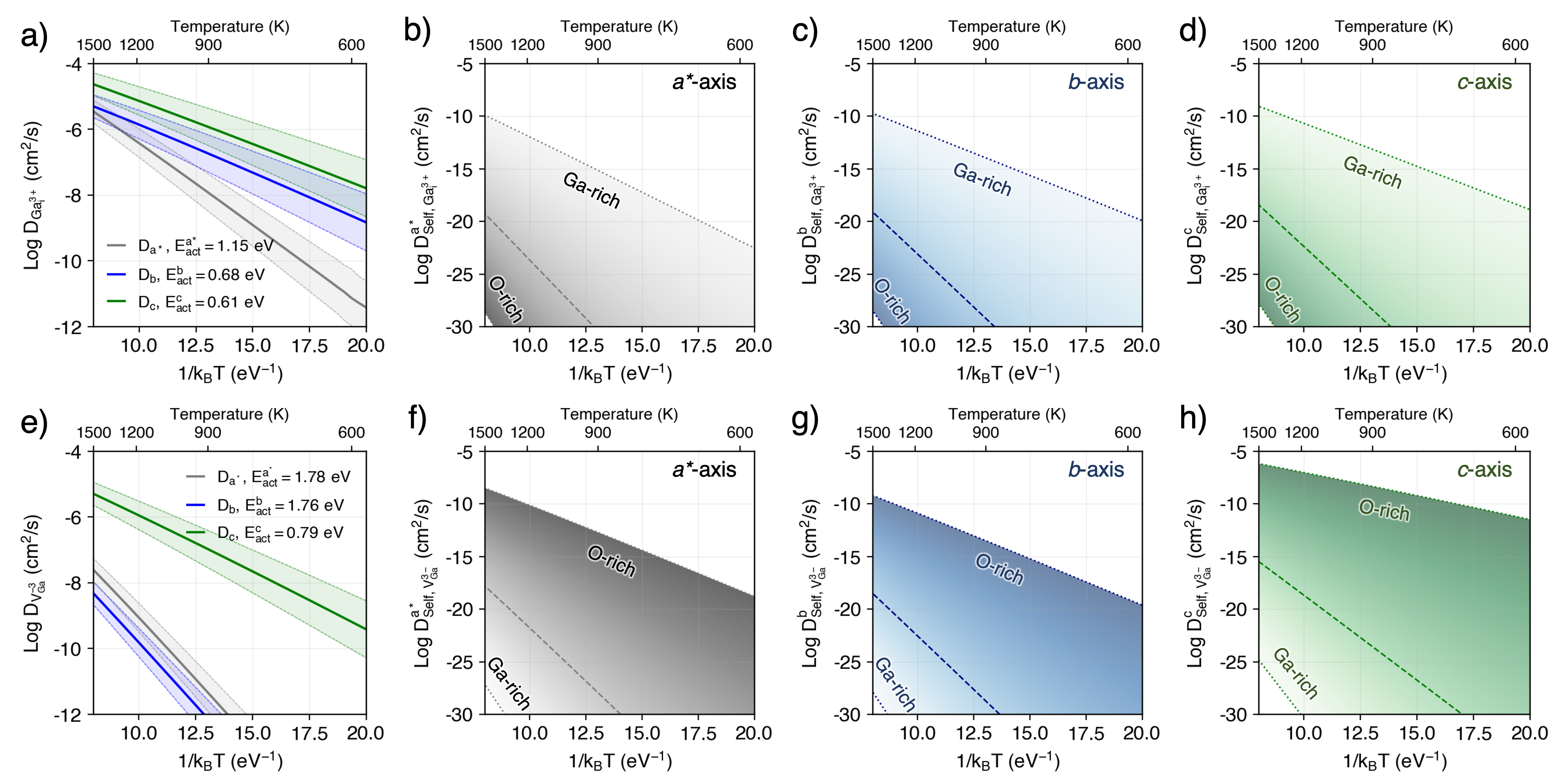}
\caption{
Arrhenius plots of the diffusion coefficients (diagonal elements) of (a) Ga$_{\text{i}}^{3+}$ and (e) V$_{\text{Ga}}^{3-}$ along three crystallographic directions ($a^*$, $b$, $c$).
The activation energies for each diffusion mechanism and direction are indicated.
The shaded areas indicate the anticipated range of diffusivity variations due to an estimated uncertainty of 0.1 eV in migration energy barriers. 
Arrhenius plots illustrating the self-diffusion coefficients of Ga mediated by (b-d) Ga$_{\text{i}}^{3+}$ and (f-h) V$_{\text{Ga}}^{3-}$ along three different crystallographic directions ($a^*$, $b$, $c$).
The dotted lines indicate the diffusivities in the Ga- or O-rich thermodynamic limit, while the dashed line denotes the intermediate state between Ga-rich and O-rich limits (halfway between Ga-rich and O-rich limits).
}
\end{figure*}

\subsubsection{Overall diffusivities and activation energies for interstitials and vacancies, \(D_{\text{Ga}_{\text{i}}^{3+}}\) and \(D_{\text{V}_{\text{Ga}}^{3-}}\)}

To construct 3D diffusion networks for interstitials Ga${_{\text{i}}^{3+}}$ and vacancies V${_{\text{Ga}}^{3-}}$, we first identified all possible hops between defects with mid-points that lie within a distance of 4 {\AA}.
This analysis resulted in a significant number of unique hops, exceeding 200 for both interstitials and vacancies.
To reduce computational burden, and given the likelihood that most hops could be decomposed into sequences of shorter substituent hops, we established a set of criteria to select a subset for first-principles NEB characterization, aiming to identify unique principle hops (PHs) that cannot be further decomposed.

For interstitial diffusion, we evaluated two key factors: (1) the number of ions undergoing large total displacements, and (2) the cumulative total displacement of all ions,  between the initial and final configurations. 
These thresholds were first evaluated assuming simple linear ion movements connecting the starting and final configurations. 
Then, we performed initial NEB calculations using 1$\times$4$\times$2 supercells for two subgroups of hops based on: (1) hops for which no ions exceed a 1.5 {\AA} threshold for total displacement, and (2) hops for which only one ion exceeded the displacement threshold while ensuring that the total summed ionic shifts remained below 10 {\AA}, as illustrated in Figure S7.
During the NEB simulations, if new metastable structures were encountered along the migration energy landscape, we isolated them and performed structural optimization to understand how the corresponding hop could be decomposed into smaller hops, as summarized in Table S1.
After all initial NEB calculations, hops that exhibited a single energy barrier and could not be further decomposed were identified as PHs of Ga${_{\text{i}}^{3+}}$.
Secondary NEB calculations were conducted for the identified PHs using larger supercells to mitigate finite-size effects. 
Supercells selected for these calculations were chosen from 1$\times$4$\times$3, 2$\times$4$\times$2, and 2$\times$4$\times$3. 
The resulting NEB migration barriers for  Ga${_{\text{i}}^{3+}}$ PHs are summarized in Table S2.
Ultimately, we ended up with 32 unique hops for Ga interstitials. 

For vacancy diffusion, we introduced an assumption that $N$-split vacancy diffusion always occurs via the formation of a ($N$-1)-split vacancy during the hop.
This assumption is made to efficiently identify the PHs by avoiding concerted movements of ions, as those movements can always be decomposed into sequences of isolated atomic hops.
For instance, a 4-split vacancy will transition to one of the 3-split vacancies first, rather than directly transforming into another 4-split vacancy. 
After applying this assumption, we performed NEB calculations specifically for those hops where the total net ionic shift did not exceed 5 {\AA}.
The resulting NEB migration barriers for  V${_{\text{Ga}}^{3-}}$ hops (31 in total) are summarized in Table S3. 

Once the diffusion pathways and their migration energy barriers were obtained from NEB calculations, the Onsager formalism \cite{42_Onsager_1944} was utilized to obtain three-dimensional diffusivity tensors for Ga${_{\text{i}}^{3+}}$ and V${_{\text{Ga}}^{3-}}$. 
Arrhenius plots for the $a^\ast$, $b$, and $c$ axis components are shown in Figure 4(a,e) for interstitials and vacancies, respectively. 
The shaded region depicts the range of diffusivity variations associated with a $\pm$0.1 eV uncertainty in calculated migration barriers.
For interstitial diffusion, the diffusion coefficients show the highest values along the $c$-axis, followed by the $b$ and $a^*$-axes.
For instance, at 600 K, the interstitial diffusion coefficients along the $a^*$, $b$, and $c$-axes are 7.42 $\times$ 10$^{-12}$, 2.34 $\times$ 10$^{-9}$, and 2.43 $\times$ 10$^{-8}$ cm$^2$/s, respectively.
For vacancy diffusion, the diffusion coefficients are largest along the $c$-axis, followed by much slower diffusion along both the $a^*$ and $b$-axes.
At 600 K, the vacancy diffusion coefficients along the $a^*$, $b$, and $c$-axes are 6.19 $\times$ 10$^{-17}$, 9.07 $\times$ 10$^{-18}$, and 6.61 $\times$ 10$^{-10}$ cm$^2$/s, respectively.
The anisotropy for vacancy diffusion is more pronounced than for interstitial diffusion, with the $c$-axis diffusion coefficient more than $10^7$ times larger than along the other axes.

The corresponding effective activation energies in Figure 4(a,e) are 1.15, 0.68, and 0.61 eV for Ga${_{\text{i}}^{3+}}$ diffusion, and 1.76, 1.78, and 0.79 eV for V${_{\text{Ga}}^{3-}}$ diffusion along the $a^*$, $b$, and $c$ --axes respectively.
The anisotropy observed for both interstitials and vacancies can potentially influence a range of key material properties for $\beta$-Ga$_2$O$_3$. 
Anisotropic diffusion of vacancies or interstitials, as well as the implied anisotropic diffusion of Ga cations (self-diffusion) and/or extrinsic dopants, could have implications for devices and thermally-activated degradation. 
For example, non-uniform diffusion with slow and fast directions could result in built-in fields that alter device performance.  
Many approaches to extrinsic doping such as ion implanation rely on diffusion, so the predicted anisotropy becomes particularly significant in situations where the diffusion of extrinsic cation dopants is mediated by vacancies or interstitials. 

\subsubsection{Ga self-diffusion and comparison to experiment}

With the vacancy and interstitial diffusivities given above, it is possible to estimate the Ga cation self-diffusivity as well. 
Self-diffusion refers to the process in which host atoms migrate through the host lattice. 
Since self-diffusion is obtained as an average over all sites including mobile and immobile ions, the Ga self-diffusion coefficient is given by the product of the fractional concentration of Ga defects (here, Ga${_{\text{i}}}$ or V${_{\text{Ga}}}$) and their respective defect diffusion coefficients \cite{51_Kabir_2020,52_Kasap_2017}:
\begin{equation}
D_{\text{self}, \text{Defect}} = \frac{C_\text{{Defect}}}{C_{\text{Ga}}} D_{\text{Defect}} \hspace{0.5em}, 
\end{equation}
where $\text{Defect} = \text{Ga}_{\text{i}},\text{V}_{\text{Ga}}$. 
This expression arises from consideration of jump balance. 
For example, the movement of Ga ions can occur via vacancy-mediated hopping: a Ga ion can jump to a neighboring Ga site if the neighboring site contains a vacancy.
After the jump, the original site becomes vacant. 
Therefore the Ga ion hop is equivalent to a V$_{\rm Ga}$ hop to an adjacent occupied Ga site taking place in the opposite direction. 
In Equation (2), $C_{\text{Defect}}$ represents the concentration of migrating Ga defects, $C_{\text{Ga}}$ denotes the total concentration of Ga sites in the bulk, and $D_{\text{Defect}}$ is the diffusivity of the defect, obtained above.  
This equation also shows that, when defect concentrations are given by equilibrium, the activation energies for self-diffusion differ from those of isolated vacancies or interstitials. 
The difference is due to the term $C_\text{{Defect}}$, which introduces an additional contribution given by the defect formation energy, due to the thermally activated nature of the defect concentration.  

\begin{figure*}[htbp!]
\centering
\includegraphics[width=6.75in]{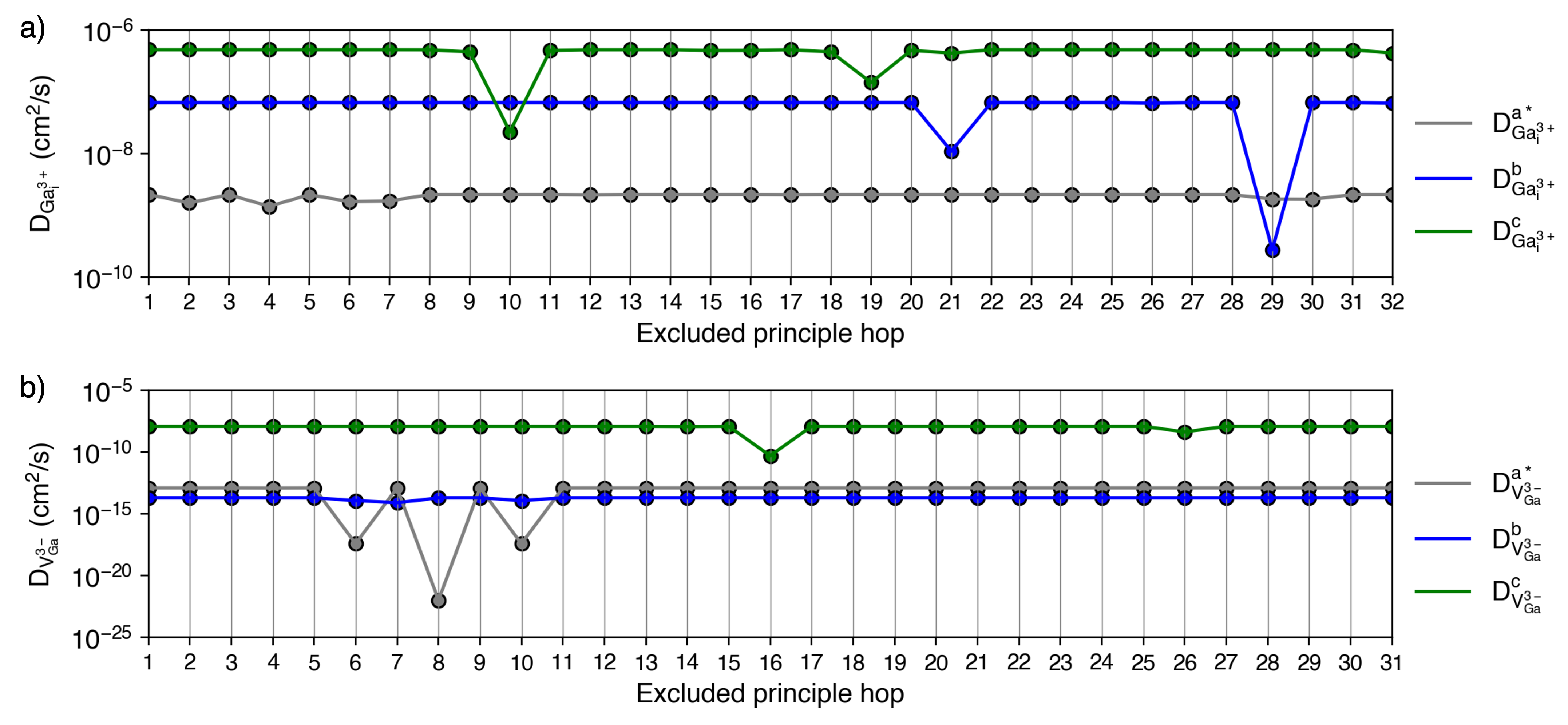}
\caption{
By excluding each principal hop one-by-one in the diffusion calculations, the reduction of the diffusion coefficients of (c) Ga$_{\text{i}}^{3+}$ and (d) V$_{\text{Ga}}^{3-}$ at $T = 800 K$ are indicated.
}
\end{figure*}

From Equation (2), were the concentrations of Ga interstitials and vacancies the same, then the slightly higher diffusivity of Ga$_{\text{i}}^{3+}$ compared to V$_{\text{Ga}}^{3-}$ (Figure 4(a,e)) would imply that the self-diffusion of Ga to be mostly mediated by Ga interstitials. 
However, we find that the substantially higher vacancy concentration expected under common n-type doping or unintentionally doped (UID) conditions (as shown in Figure S6) is sufficient to make Ga self-diffusion entirely dominated by vacancies. 
Figure 4(b-d) and 4(f-h) present Arrhenius plots for the self-diffusivities mediated by Ga interstitials ($\text{Ga}_{\text{i}}^{3+}$) and Ga vacancies ($\text{V}_{\text{Ga}}^{3-}$) across an wide spectrum of thermodynamic conditions, ranging from Ga-rich (low $\text{P(O}_2$)) to O-rich (high $\text{P(O}_2$)), respectively.  
These diffusivities are obtained under the assumption that the Fermi level is positioned at 3.0 eV (1.8 eV below the conduction band minimum (CBM)). 
In an intermediate condition between Ga-rich and O-rich limits, depicted by dashed lines, self-diffusion coefficients for $\text{V}_{\text{Ga}}^{3-}$ are predicted to be higher across all crystallographic axes, attributed to significant differences in equilibrium defect concentrations. 
For example, $\text{D}_{\text{self}, \text{V}_{\text{Ga}}^{3-}} = 7.00 \times 10^{-19}$ cm$^2$/s but  $\text{D}_{\text{self}, \text{Ga}_{\text{i}}^{3+}} = 1.87 \times 10^{-22}$ cm$^2$/s along the $c$-axis at 1200 K.
In the O-rich limit, applicable to diffusion experiments at ambient pressures, the difference in Ga self-diffusion coefficients becomes more significant. 
Now, $\text{D}_{\text{self}, \text{V}_{\text{Ga}}^{3-}} = 1.33 \times 10^{-7}$ cm$^2$/s and $\text{D}_{\text{self}, \text{Ga}_{\text{i}}^{3+}} = 8.95 \times 10^{-34}$ cm$^2$/s along the $c$-axis at 1200 K. 

\subsubsection{Comparison to experimental measurements of activation energies for defect migration}

Experimental measurements of the activation energy for ionic diffusion or ionic conductivity are challenging, since multiple ionic transport processes can contribute to measured quantities. It is often difficult to know \textit{a priori} which processes dominate.  
The simultaneous contribution of electrons to measured conductivities adds to the complexity of isolating ionic contributions alone.
In semi-insulating or UID $\beta$-Ga$_{2}$O$_{3}$, where electronic contributions are largely suppressed, the diffusion of charged defects such as compensating V$_{\text{O}}^{2+}$, Ga$_{\text{i}}^{3+}$ and V$_{\text{Ga}}^{3-}$ becomes the dominant contributor to charge transport.
In such cases, the measured activation energy of carrier mobility or of conductivity is a reasonable estimate of the activation energy for charged defect diffusion.

Fleischer et al.\ reported activation energies for carrier mobility of 0.60 eV and 0.65 eV for UID crystal and ceramics, respectively \cite{45_Fleischer_1993}.
Ghadbeigi et al.\ reported an activation energy of 0.86 eV for total conductivity in semi-insulating Mg-doped $\beta$-Ga$_{2}$O$_{3}$ under comparable contributions from electronic and ionic conduction \cite{46_Ghadbeigi_2022}.
As suggested by Kyrtsos et al., migration barriers for V$_{\text{O}}^{2+}$ are approximately 1 eV higher than for Ga$_{\text{i}}^{3+}$ and V$_{\text{Ga}}^{3-}$  \cite{13_Kyrtsos_2017}, so these experimentally reported activation energies are likely attributed to the migration of charged Ga defects. 
Also, Ingebrigtsen et al.\ experimentally measured an activation energy of 1.2 eV using electrical conductivity recovery measurements \cite{17_Ingebrigtsen_2018} that they associated with gallium vacancy migration.  
However, these energies were obtained in proton implanted $\beta$-Ga$_2$O$_3$ in the context of the recovery of electronic charge carrier concentrations under subsequent annealing. 
More recently, Azarov et al.\ reported a migration barrier of 0.80 eV, determined by a dose-rate effect methodology, which assesses the impact of ion flux and temperature on lattice disorder and defect migration \cite{47_Azarov_2021}. 

Overall, our calculated lowest migration barriers for Ga$_{\text{i}}^{3+}$ and V$_{\text{Ga}}^{3-}$ agree well with the findings of Fleischer et al., Ghadbeigi et al., and Azarov et al., but not with the barrier of 1.2 eV reported by Ingebrigtsen et al.
Although the reason for the discrepancy with Ingebrigtsen's result is not known, we suggest that it could be attributed to their measurement reflecting second-order kinetic processes in addition to vacancy migration during the thermal recovery process. 
Under second-order kinetics, processes such as diffusion followed by subsequent defect trapping following reaction equations can introduce additional contributions to activation energies beyond the pure defect migration barriers.

\begin{figure*}[htbp!]
\centering
\includegraphics[width=6.25in]{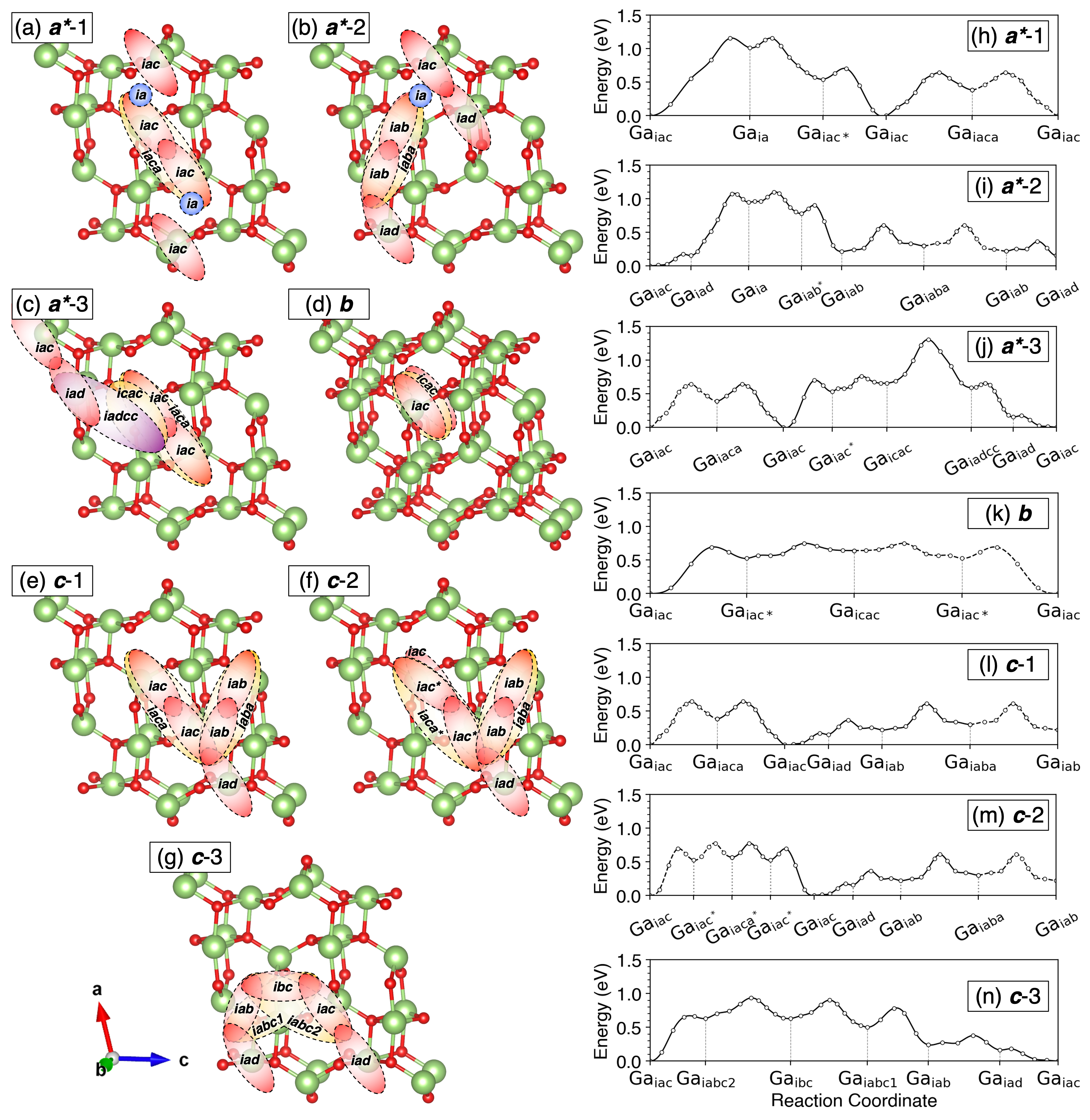}
\caption{
Schematic illustrations of dominant diffusion pathways of Ga$_{\text{i}}^{3+}$ along (a-c) $a^*$-axis, (d) $b$-axis, and (e-g) $c$-axis crystal orientations, respectively. 
(h-n) Corresponding energy landscapes along the energy minimum pathways.
Overlapping principal interstitial hops within the diffusion pathways are represented by dashed lines in the energy profiles.
}
\end{figure*}

\subsubsection{Dominant migration pathways for interstitial diffusion}

The diffusion coefficients and effective activation energies reported in Figure 4(a,e) arise from the complete diffusion network, comprising all possible defect sites and all possible hops between them. 
These coefficients arise from multiple contributions. 
First, under equilibrium, the site probabilities $\rho_i$ (see Appendix A) are described by Boltzmann statistics. 
The overall residence time for which a defect occupies a higher energy configuration is smaller than its residence time in lower energy configurations. 
Consequently, if diffusion along a particular crystallographic direction necessitates passing through a site with large energy (relative to other available sites), then the need to pass through this site introduces diffusion bottlenecks. 
Second, even for transitions between two comparatively low-energy sites, the transition barrier through which the defect passes may be large, also slowing the diffusion. 
Under equilibrium (see Appendix A), transition rates obey detailed balance so that $\rho_i \lambda_{i \rightarrow j} = \rho_j \lambda_{j \rightarrow i}$ where $\lambda_{i \rightarrow j}$ is the transition rate from site $i$ to site $j$ and is proportional to $\exp{(-\beta 
[E^{ts}_{ij} - E_i])}$. 

To identify the reasons underlying the differences in diffusivities along different crystallographic axes (Figure 4(a,e)), it is necessary to first  
isolate the contribution of each PH and identify the sites and hops that are dominant contributors to diffusion in each crystallographic direction. 
For this, we constructed reduced diffusion networks by excluding one PH at a time from the full set of principal hops for Ga${_{\text{i}}^{3+}}$ diffusion. 
As shown in Figure 5, by comparing the resulting diffusivities with those obtained from the full set described in the previous section,  we were able to assess the contribution of each PH to the total diffusivity. 
For example, when PH 10 or PH 19 is removed from the diffusion network of Ga${_{\text{i}}^{3+}}$ (Figure 5(b)), a substantial decrease in diffusivity along the $c$-axis is observed, indicating the critical role of these two hops in facilitating Ga${_{\text{i}}^{3+}}$ diffusion along the $c$-axis. 
After identifying dominant contributors, we constructed additional reduced diffusion networks that included only these core PHs and compared its diffusivity with that of the full set, allowing us to isolate the set of active diffusion pathways formed by combinations of core PHs.
Through this analysis, we determined the dominant diffusion pathways for Ga${_{\text{i}}^{3+}}$ diffusion along each crystallographic direction.  These pathways and the corresponding energy landscapes are illustrated in Figure 6. 

For Ga${_{\text{i}}^{3+}}$ diffusion along the $a^*$-axis (the slowest axis), we aimed to identify the main bottlenecks. We identified three major hopping pathways which are shown in Figure 6(a-c). 
The pathways shown in Figure 6(a,b) are associated with $ac$-plane diffusion, following the Ga${_{\text{iac}}}$- Ga${_{\text{iaca}}}$-Ga${_{\text{iac}}}$ and Ga${_{\text{iab}}}$- Ga${_{\text{iaba}}}$-Ga${_{\text{iab}}}$ routes shown in Figure 6(e). 
The associated energy landscapes are shown in Figure 6(h,i). 
The landscapes show that diffusion along $a^\ast$ requires passing through the high energy Ga$_{\text{ia}}$ site, which is 1.01 eV higher in energy than the lowest energy Ga$_{\text{iac}}$ site. 
For both of these paths, we observe similar rate-limiting steps in which a favorable split interstitial (Ga${_{\text{iab}}}$ or Ga${_{\text{iad}}}$) diffuses across the hexagonal channel along the $b$-axis to a high-energy Ga${_{\text{ia}}}$ configuration with migration energies of approximately 1.16 and 0.92 eV. 
In the alternative $a^\ast$ pathway of Figure 6(c,j), a slightly more complex route, the quadruple split interstitial Ga${_{\text{iadcc}}}$ is employed to bypass the high-energy Ga${_{\text{ia}}}$ state.  Unfortunately, this route suffers from a large transition barrier however, 0.72 eV between Ga${_{\text{iadcc}}}$ and Ga${_{\text{icac}}}$. 
By comparing the diffusivity of reduced diffusion networks that include each set of selected PHs in Figure 6(a-c), we obtained $a^*$-axis diffusivities of 38\%, 44\%, and 16\% from the total $a^*$-axis diffusivity, respectively.
When including all hops in Figure(a-c), we achieve 98\% of the diffusivity, indicating that these three paths make up the main diffusion pathways along the $a^\ast$ axis. 

\begin{figure*}[htbp!]
\centering
\includegraphics[width=6.25in]{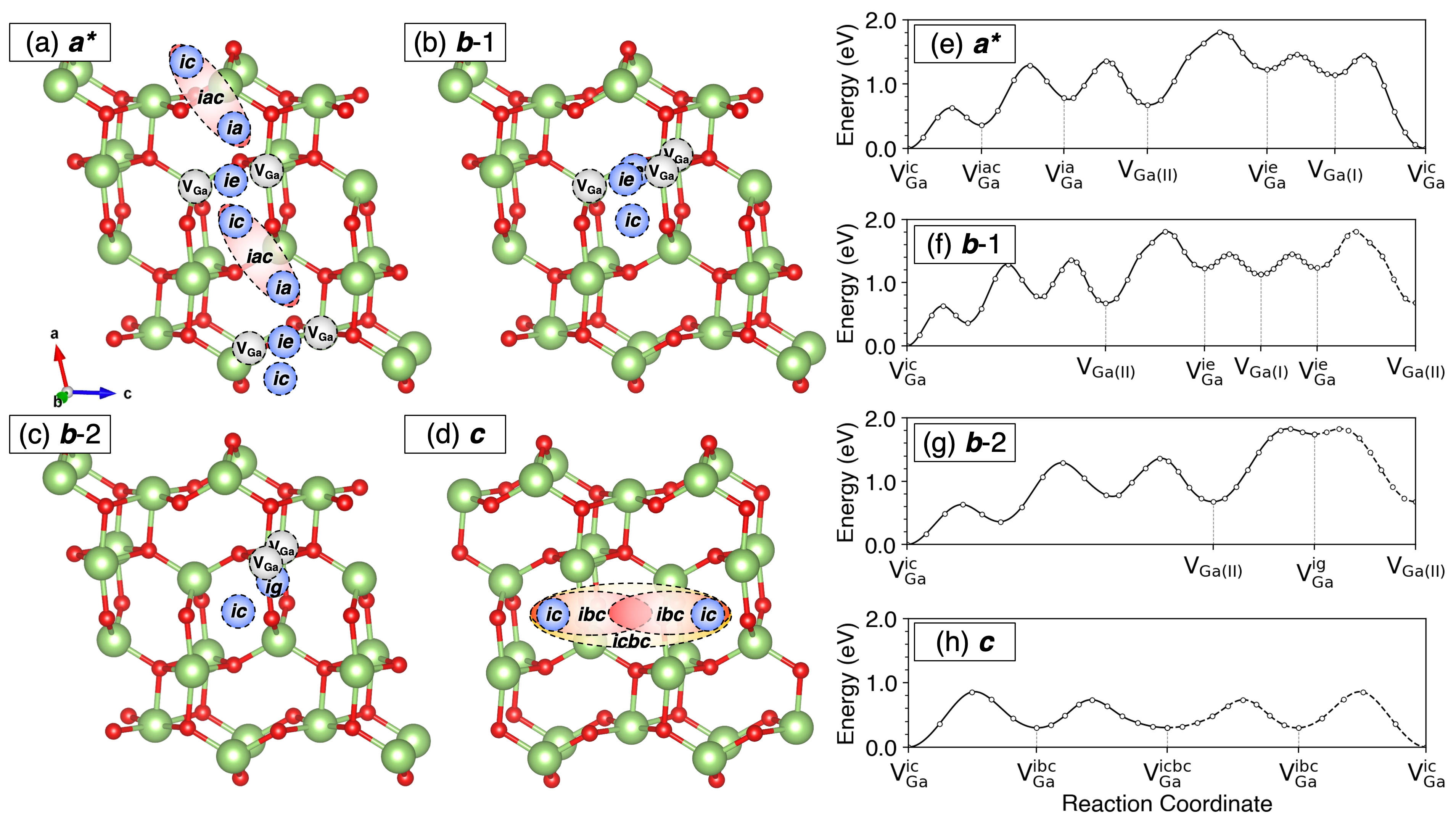}
\caption{
Schematic illustrations of dominant diffusion pathways of V$_{\text{Ga}}^{3-}$ along (a) $a^*$-axis, (b,c) $b$-axis, and (d) $c$-axis crystal orientations, respectively. 
(e-h) Corresponding energy landscapes along the energy minimum pathways.
}
\end{figure*}

Regarding $b$-axis diffusion, the direction with intermediate diffusivity, 
we identified one major diffusion pathway (Figure 6(d,k)).  
In this pathway, the Ga${_{\text{iac}}}$ interstitial moves across the hexagonal channel along the $b$-axis and forms a symmetric three-split interstitial, Ga${_{\text{icac}}}$. 
Unlike the $a^\ast$ direction, there is no need to pass through high energy sites like Ga$_{\text{ia}}$, the interstitial remains as a split throughout the entire path, and the energy profile in Figure 6(k) is smooth compared to the paths in Figure 6(h,i,j). 
The highest site energy of 0.64 eV is associated with Ga${_{\text{icac}}}$. 
The rate-limiting step occurs during the Ga${_{\text{iac}}}$-Ga${_{\text{iac}^*}}$ process with transition barrier of 0.69 eV, where the split interstitial deviates from the $ac$-plane along with the other $ic$ Ga split interstitial and two Ga vacancies, resulting in the loss of $b$-axis symmetry.  
This hopping pathway accounts for over 97\% of the diffusivity along the $b$-axis.

For $c$-axis diffusion (the fastest direction), we identified three contributing diffusion pathways. 
The pathways are illustrated in Figure 6(e,f,g), and the potential energy surfaces are shown in Figure 6(l,m,n). 
Similar to diffusion along the $a^*$-axis, fast in-plane diffusion pathways are observed following the Ga${_{\text{iab}}}$-Ga${_{\text{iaba}}}$-Ga${_{\text{iab}}}$ and Ga${_{\text{iac}}}$-Ga${_{\text{iaca}}}$-Ga${_{\text{iac}}}$ routes, as shown in Figure 6(e,l). 
In the pathway of Figure 6(f,m), the Ga${_{\text{iac}}}$-Ga${_{\text{iac}^*}}$ process is involved, resulting in another side path, Ga${_{\text{iac}^*}}$-Ga${_{\text{iaca}^*}}$-Ga${_{\text{iac}^*}}$, with moderate migration barriers. 
In the pathway Figure 6(g,n), the Ga${_{\text{ibc}}}$ site is utilized to connect the 3-split interstitials of Ga${_{\text{iabc1}}}$ and Ga${_{\text{iabc2}}}$.
By comparing the diffusivity of reduced diffusion networks these three paths, we obtained diffusivities of 84\%, 26\%, and 3\% of the $c$-axis total diffusivity, respectively. 
Considering all hops in Figure 6(e-g), we achieve 99\% of the diffusivity. 
Since all $c$-axis diffusion pathways occur predominantly within the $ac$-plane, the migration barriers are smaller compared to the other two directions, resulting in higher diffusivities along $c$ compared to $a^\ast$ and $b$.

Overall, we attribute the higher diffusivity along $c$ compared to $b$ to the availability of multiple paths, some of which show slightly lower barriers than the single path available for $b$-axis diffusion.
Compared to the $b$ direction, the three $c$-axis pathways involve sites of similar site energies.
For path $c$-1, we have Ga$_{\text{iaca}}$ (0.39 eV) and Ga$_{\text{iaba}}$  (0.30 eV). 
For path $c$-2, we have Ga$_{\text{iac}^{\ast}}$ (0.53 eV) and Ga$_{\text{iaca}^{\ast}}$  (0.57 eV).
And for path $c$-3, we have Ga$_{\text{iabc2}}$ (0.63 eV), Ga$_{\text{ibc}}$ (0.63 eV), and Ga$_{\text{iabc1}}$ (0.50 eV).
The associated rate limiting barriers for each path are 0.64, 0.69, and 0.66 eV respectively. 

\subsubsection{Dominant migration pathways for vacancy diffusion}

We also explored the dominant pathways for V${_{\text{Ga}}^{3-}}$ diffusion along each crystallographic direction.
The dominant pathways and their corresponding energy landscapes are illustrated in Figure 7.
Along the $a^*$-axis (a very slow direction), we identified one predominant pathway, shown in Figure 7(a,e). 
The landscape indicates that diffusion along $a^\ast$ necessitates passing through several high-energy states, including V${_{\text{Ga(I)}}}$ and V${_{\text{Ga}}^{\text{ie}}}$, with site energies that are 1.14 eV and 1.22 eV above the lowest energy V${_{\text{Ga}}^{\text{ic}}}$ state, respectively.
The path starts with elementary hops between the lowest-energy 2-split V${_{\text{Ga}}^{\text{ic}}}$ and the 3-split V${_{\text{Ga}}^{\text{iac}}}$, 
which occurs via a Ga(II) ion moving into the $ia$ interstitial site. 
In the adjacent unit cell, the $ic$ split interstitial (present as part of V${_{\text{Ga}}^{\text{iac}}}$) 
moves to the Ga(I) site to form V${_{\text{Ga}}^{\text{ia}}}$. 
Next, the $ia$ split interstitial 
(present as part of V${_{\text{Ga}}^{\text{ia}}}$)  moves to the Ga(I) site, leaving V${_{\text{Ga(II)}}}$. 
Then, Ga(I) moves into the $ie$ split intersititial site,  forming 2-split V${_{\text{Ga}}^{\text{ie}}}$. 
Finally, the $ie$ split interstitial (present as part of V${_{\text{Ga}}^{\text{ie}}}$) moves to Ga(II).
The rate-limiting step, with a migration barrier of 1.13 eV, occurs during the V${_{\text{Ga(II)}}}$-V${_{\text{Ga}}^{\text{ie}}}$-V${_{\text{Ga(I)}}}$ process, involving ionic movements along the $b$-axis. 
This hopping pathway accounts for over 99\% of the diffusivity along the $a^*$-axis.

For $b$-axis diffusion (the slowest direction), we identified two major pathways. These are illustrated in Figure 7(b,c) and their energy landscapes are given in Figure 7(f,g).
Similar to the $a^{\ast}$, the diffusion path $b$-1 necessitates passing through high energy states of V${_{\text{Ga(I)}}}$ and V${_{\text{Ga}}^{\text{ie}}}$.
This path follows the V${_{\text{Ga(II)}}}$-V${_{\text{Ga}}^{\text{ie}}}$-V${_{\text{Ga(I)}}}$ process similar to Figure 7(a,e), and shows a rate-limiting barrier of 1.13 eV.
The second path in Figure 7(c,g) proceeds directly along the $b$-axis by passing through V${_{\text{Ga}}^{\text{ig}}}$, with high site energy of 1.74 eV. 
It follows the V${_{\text{Ga(II)}}}$-V${_{\text{Ga}}^{\text{ig}}}$-V${_{\text{Ga(II)}}}$ route; the rate-limiting barrier of 1.15 eV occurs when passing through V${_{\text{Ga}}^{\text{ig}}}$.
By comparing the diffusivity of reduced diffusion networks, we obtain diffusivities of 60\% and 40\% of the $b$-axis diffusivity, respectively. 
Considering all hops in Figure 7(b,c), we achieve 99\% of the total diffusivity. 

In the case of $c$-axis diffusion, substantially faster than the other two directions,  one major pathway is identified, depicted in Figure 7(d,h). 
Unlike diffusion along the $a^\ast$ and $b$-axis, the vacancy moves directly along the $c$-axis by passing through low-energy states of V${_{\text{Ga}}^{\text{ibc}}}$ and V${_{\text{Ga}}^{\text{icbc}}}$, with site energies for both of only 0.30 eV, thereby avoiding high-energy states.
This pathway involves splitting of the lowest-energy 2-split V${_{\text{Ga}}^{\text{ic}}}$ into the 3-split V${_{\text{Ga}}^{\text{ibc}}}$. 
Instead of the V${_{\text{Ga}}^{\text{ic}}}$-V${_{\text{Ga}}^{\text{ibc}}}$-V${_{\text{Ga}}^{\text{ic}}}$ route proposed by Frodason et al.\ \cite{16_Frodason_2023}, the presence of the additionally extended 4-split V${_{\text{Ga}}^{\text{icbc}}}$ allows bypassing the V${_{\text{Ga}}^{\text{ic}}}$-V${_{\text{Ga}}^{\text{ibc}}}$ energy barrier, resulting in a different  V${_{\text{Ga}}^{\text{ic}}}$-V${_{\text{Ga}}^{\text{ibc}}}$-V${_{\text{Ga}}^{\text{icbc}}}$-V${_{\text{Ga}}^{\text{ibc}}}$-V${_{\text{Ga}}^{\text{ic}}}$ route.
The rate-limiting step occurs along V${_{\text{Ga}}^{\text{ic}}}$-V${_{\text{Ga}}^{\text{ibc}}}$,  with a barrier of 0.86 eV.
This hopping pathway accounts for over 99$\%$ of the diffusivity along the $c$-axis.
Here, the identification of the 4-split V${_{\text{Ga}}^{\text{icbc}}}$ creating the lowest energy pathway highlights the potential for exploiting further extended $N$-split vacancies (N $\geq$ 4) to access new low-energy diffusion paths and suggests that further exploration of more extended $N$-split defects on larger scales could be fruitful.

\subsubsection{Comparison to results of Frodason \textit{et al.}}

Overall, our predictions for the dominant diffusion pathways are in large agreement with those proposed by Frodason \cite{16_Frodason_2023}, showing only minor differences. 

Regarding the diffusion of Ga$_{\text{i}}^{3+}$, we predict the fastest diffusion along $c$, followed by $b$, and then $a^*$. 
In contrast, Frodason et al.\ suggest the order to be $a^*$, $c$, and $b$-axes. 
This discrepancy can be attributed to the reduced energy barriers for both Ga$_{\text{iaba}}$ and Ga${_{\text{iaca}}}$ transition states obtained here, which are critical for diffusion along the $c$-axis.
This reduction was observed after employing larger 2$\times$4$\times$2 supercells in our simulations. 
For $c$-axis diffusion, we predict the fastest pathway as Ga${_{\text{iac}}}$-Ga${_{\text{iad}}}$-Ga${_{\text{iab}}}$, while  Frodason et al. suggest the Ga${_{\text{iac}}}$-Ga${_{\text{iab}}}$ route.
Both studies agree on the diffusion route along the $b$-axis. 
Lastly, for $a^*$-axis diffusion, we predict the fastest route along in the Ga${_{\text{iac}}}$-Ga${_{\text{iac}^*}}$-Ga${_{\text{ia}}}$-Ga${_{\text{iac}}}$ route, while Frodason et al.\ suggest the Ga${_{\text{iac}}}$-Ga${_{\text{iac}^*}}$-Ga${_{\text{if}}}$-Ga${_{\text{ia}}}$-Ga${_{\text{iac}}}$ route. 

Regarding the V${_{\text{Ga}}^{3-}}$ diffusion pathways, both studies agree on the order of fast diffusion directions as $c$, $a^*$, and $b$-axes.
Both studies also predict the same routes along the $a^*$ and $b$ axes. 
However, we introduce a slightly modified route for $c$-axis diffusion by incorporating the 3-split V${_{\text{Ga}}^{\text{icbc}}}$ structure.

Despite minor differences, the overall agreement between two independent studies gives credence to the findings and highlights the importance of $N$-split defects in understanding the diffusion mechanisms of $\beta$-Ga$_2$O$_3$.

\section{Conclusion}

We used first-principles calculations to explore various Ga native defect configurations in $\beta$-Ga$_2$O$_3$, including $N$-split defects, and analyzed operative vacancy and interstitial diffusion networks.
By employing the Onsager approach, we formulate the master diffusion equations for Ga interstitials and Ga vacancies by constructing a 3D diffusion network from a comprehensive set of 32 unique interstitial and interstitialcy hops between 20 different configurations of Ga interstitials and 31 unique vacancy hops between 19 different configurations of Ga vacancies.
The solution of these equations yields the three-dimensional diffusivity tensors.
Both Ga interstitials and vacancies demonstrated the highest diffusivity along the $c$-axis with calculated diffusivities of 8.89$\times$10$^{-6}$ and 1.45$\times$10$^{-6}$ cm$^2$/s at $T = 1200$ K, corresponding to the lowest effective activation energies of 0.61 eV and 0.79 eV, respectively. 
However, Ga self-diffusion is ultimately mediated by vacancies, because of the substantially higher concentration of Ga vacancies than Ga interstitials; For instance, we estimate self-diffusion coefficients of $1.33 \times 10^{-7}$ and $8.95\times 10^{-34} \, \text{cm}^2/\text{s}$ for processes mediated by $\text{Ga}_\text{i}^{3+}$ and $\text{V}_\text{Ga}^{3-}$ at $T = 1200$ K under the O-rich condition, respectively.
We also identified the dominant diffusion mechanism in each crystallographic direction. 
We hope these findings improve the understanding of intrinsic defect diffusion in $\beta$-Ga$_2$O$_3$, and help in understanding degradation and other mass transport related phenomena in high-performance power devices.

\section{Acknowledgment}
The authors acknowledge the funding provided by the Air Force Office of Scientific Research under Award No. FA9550-21-0078 (Program Manager: Dr. Ali Sayir).
This work used PSC Bridges-2 at the Pittsburgh Supercomputing Center through allocation MAT220011 from the Advanced Cyberinfrastructure Coordination Ecosystem: Services \& Support (ACCESS) program, which is supported by National Science Foundation grants \#2138259, \#2138286, \#2138307, \#2137603, and \#2138296.

\section{DATA AVAILABILITY}
The data that support the findings of this study are available on GitHub at https://github.com/ertekin-research-group/2024-Ga2O3-Ga-Diffusion, Ref. 51.\cite{2024-Ga2O3-Ga_Diffusion}

\section{Appendix}

\subsection{Direct Onsager approach: Diffusivity calculation}

The Onsager approach for obtaining diffusivity tensors is based on several key assumptions. 
First, it is assumed that defects undergo harmonic motion within the crystal lattice, allowing for simplification of the master diffusion equations to describe the time evolution of the system. 
Second, it is assumed that the equilibrium site probabilities $\rho_i$ follow a Boltzmann  relationship amongst thermally occupied energy states, given by 
\[ \rho_i = \frac{1}{Z} \rho_{i}^0 \exp(-\beta E_i) \hspace{0.5em}, \]  
where $ \rho_{i}^0 = \exp\left(\frac{S_i}{k_B}\right) $ represents the entropic prefactor for the static state, estimated here to be uniformly 1 for all configurations. Here,  $Z$ is the partition function, which is defined by the sum of the Boltzmann factors over all available states as $Z = \sum_i \rho_{i}^0 \exp(-\beta E_i)$.
Third, from transition state theory, it is assumed that the transition rates $ \lambda_{i \rightarrow j} $ under dilute conditions follow 
\[ \lambda_{i \rightarrow j} = \frac{\lambda_{ij}^0}{\rho_{i}^0} \exp(-\beta[E^{ts}_{ij} - E_i]) \hspace{0.5em}, \] 
where $\lambda_{ij}^0 = \exp\left(\frac{S^{ts}_{ij}}{k_B}\right)$ represents the entropic prefactor for the transition state, i.e.\ the attempt frequency for the particular jump. 
In our work, we approximate this prefactor using a typical phonon frequency of 10$^{13}$ Hz \cite{13_Kyrtsos_2017,16_Frodason_2023}. 

With site energies and transition barriers enumerated and obtained from first-principles, 
the Onsager software package \cite{22_Trinkle_Onsager} is used to construct the master diffusion equations. The steady-state solution of these equations results in the diffusion tensor $D$, which can be expressed as 
\[ D = \frac{1}{2} \sum_{ij} \delta x_{i \rightarrow j} \otimes \delta x_{i \rightarrow j} \lambda_{i \rightarrow j} \rho_i + \sum_i b_i \otimes \gamma_i \hspace{0.5em}. \] 
Here, $\delta x_{i \rightarrow j} = x_j - x_i$ represents the displacement of the diffusing defect from state $i$ to state $j$, $\rho_i$ is the equilibrium site probability for site $i$, $b_i$ is the scaled velocity vector representing the bias of jumps at site $i$, and $\gamma_i$ is the bias-correction vector obtained by solving $\sum_j \omega_{ij} \gamma_j = b_i$, which accounts for correlations arising from unbalanced forward and backward jumps due to the varying local atomic environment.

\subsection{Defect formation energy calculation}

To obtain the defect formation energies and charge transition levels for Ga defects, we followed the standard supercell approach \cite{37_Freysoldt_2014,38_Lany_2009,39_Adamczyk_2021}. 
The formation energy $E_f[X^q]$ of a defect $X$ in charge state $q$ was obtained by determining the energy difference between the supercell containing the defect $X$ and the pristine bulk $\beta$-Ga$_2$O$_3$ supercell according to 
\[ E_{f}[X^q] = E_{\text{tot}}[X^q] - E_{\text{tot}}[\text{Bulk}] - n \mu_{\text{Ga}} + qE_{\text{Fermi}} + E_{corr} \hspace{0.5em}. \]
Here, $E_{\text{tot}}[X^q]$ and $E_{\text{tot}}[\text{Bulk}]$ represent the total energy of the supercell containing defect $X$ in charge state $q$ and the total energy of the host pristine supercell, respectively. 
The term $\mu_{\text{Ga}}$ represents the Ga chemical potential, and $n = 1$ ($n = -1$) when a Ga species is added to (removed from) the supercell to create the defect. 
The charging of defects involves the exchange of electrons with the electron chemical potential (semiconductor Fermi level, $E_{\text{Fermi}}$), typically referenced to the valence-band maximum. 

Two limits were considered as the upper and lower bounds for the Ga chemical potential. 
In the upper limit (Ga-rich), $\mu_{\text{Ga}}$ is given by the energy per Ga atom in the pure elemental Ga phase ($\mu_{\text{Ga}}^0$). 
In the lower limit (O-rich), $\mu_{\text{O}}$ is given by half of 
the energy of an O$_2$ molecule ($\mu_{\text{O}}^0$) under 1 atm 
and 1000 $K$ conditions.  
In this latter case,  $\mu_{\text{Ga}}$ is shifted from $\mu_{\text{Ga}}^0$ following the thermodynamic stability 
condition $\mu_{\text{Ga}} = \mu_{\text{Ga}}^0 + \frac{1}{2}\Delta H_f (\beta\text{-Ga}_2\text{O}_3)$, where $\Delta H_f (\beta\text{-Ga}_2\text{O}_3)$ is the calculated formation energy of $\beta$-Ga$_2$O$_3$ (10.78 eV per f.u.).

To compensate for the finite-size effects caused by electrostatic interactions between charged defects in neighboring supercells, we incorporated the energy correction term $E_{\text{corr}}$. 
We adopted the methodology proposed by Lany and Zunger \cite{38_Lany_2009} to estimate the energy corrections for potential alignment $\Delta E_{\text{pa}}(D,q)$ and image charge $\Delta E_{\text{i}}$, expressed by
\[
\Delta E_{\text{pa}}(D,q) = q \cdot \Delta V_{pa} \hspace{0.5em},
\]
\[
\Delta E_{\text{i}} = \frac{q^2\alpha_M}{2\epsilon\omega^{-1/3}} \hspace{0.5em}.
\]
Here, $\Delta E_{\text{pa}}$ represents the potential alignment between the defect and the host supercell, $\alpha$ denotes the Madelung constant specific to the supercell geometry, $\epsilon$ corresponds to the static dielectric constant, and $\omega$ represents the volume of the supercell. 

\bibliography{PRM_Main.bib}

\end{document}


\renewcommand\thefigure{S\arabic{figure}} 
\newcounter{fig2} 

\begin{figure*}[!hbtp] 
\centering
\includegraphics[width=5in]{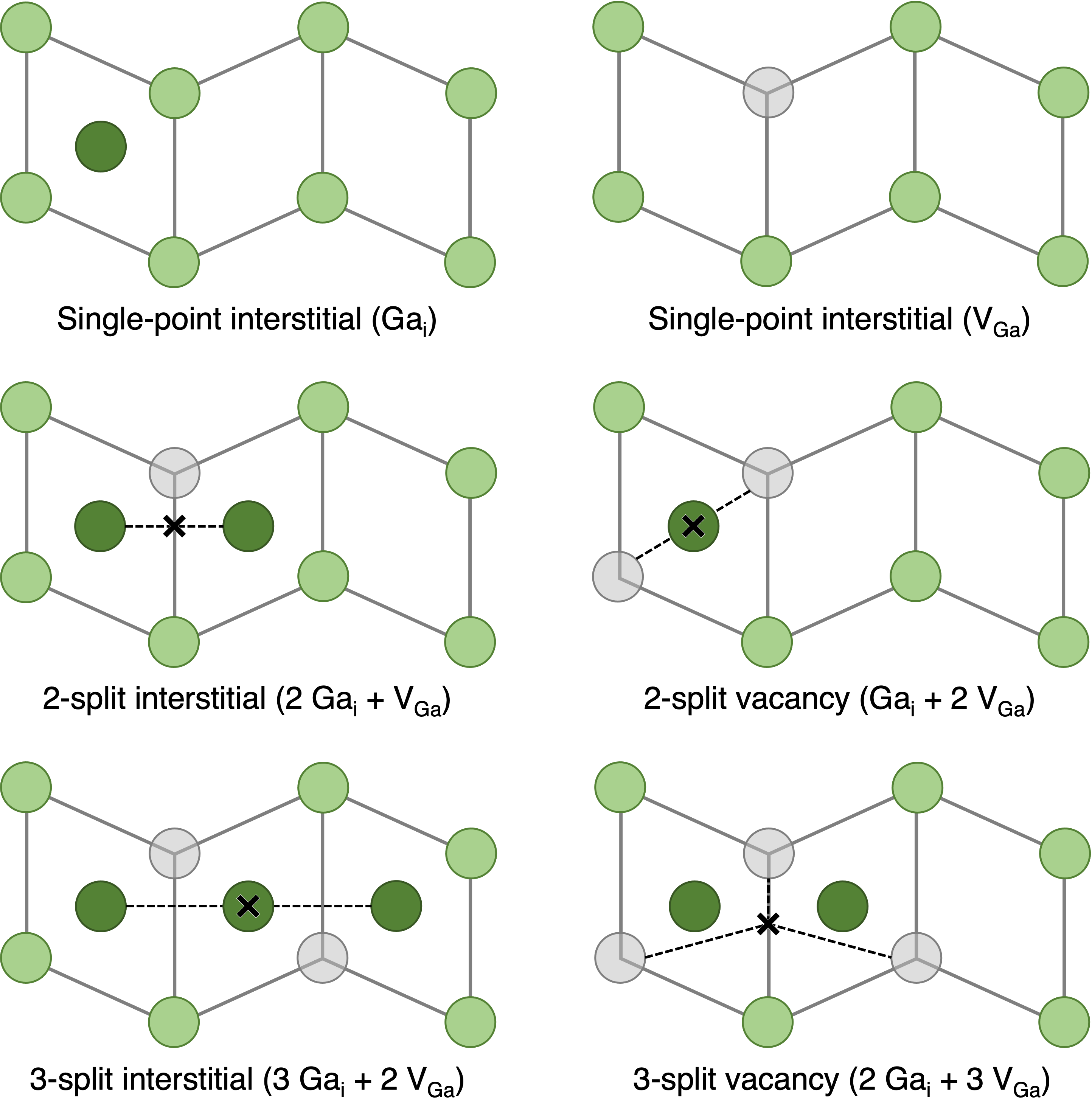}
\caption{
Schematic illustrations of the single-point Ga interstitial and the single-point vacancy, as well as $N$-split interstitials and $N$-split vacancies, with each featuring an 'X' marking the midpoint of split defects to simplify tracking of the averaged locations of scattered split defects.
The dark green atoms represent the Ga interstitials, while the gray transparent atoms represent the Ga vacancies.
}
\end{figure*}

\newpage

\begin{figure*}[!hbtp] 
\centering
\includegraphics[width=6.5in]{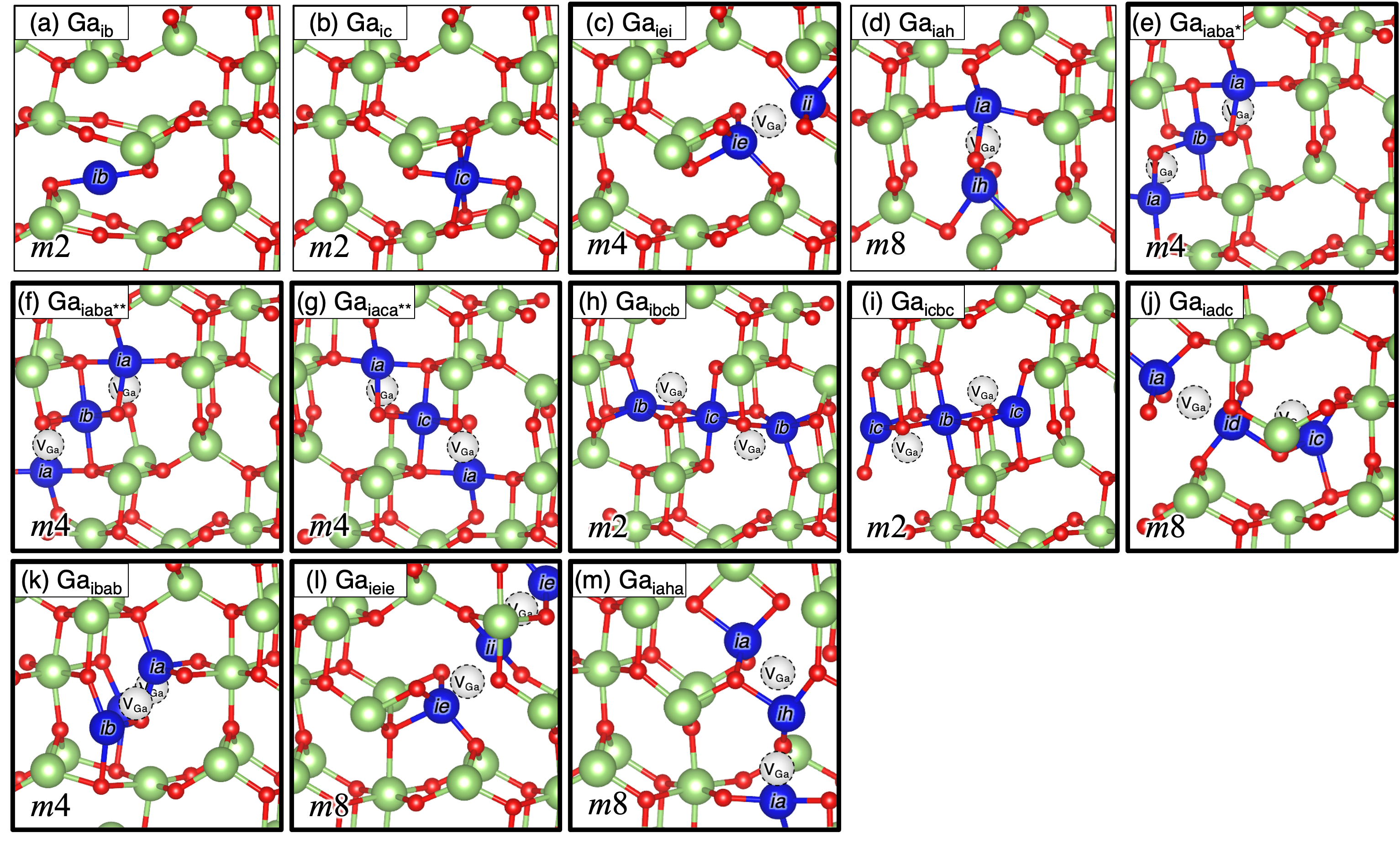}
\caption{
The relaxed $\beta$-Ga$_2$O$_3$ structures with Ga$_{\text{i}}^{3+}$, which are not included in the main text, are also categorized as (a,b) single, (c,d) double-split, and (e-m) triple-split interstitials. Ga interstitials are represented by blue balls, while Ga vacancies are depicted by gray balls with dashed circles. The Greek letter in the lower left corner denotes the multiplicity of Ga$_{\text{i}}^{3+}$ configurations within the unit cell.
}
\end{figure*}

\newpage

\begin{figure*}[!hbtp] 
\centering
\includegraphics[width=6.5in]{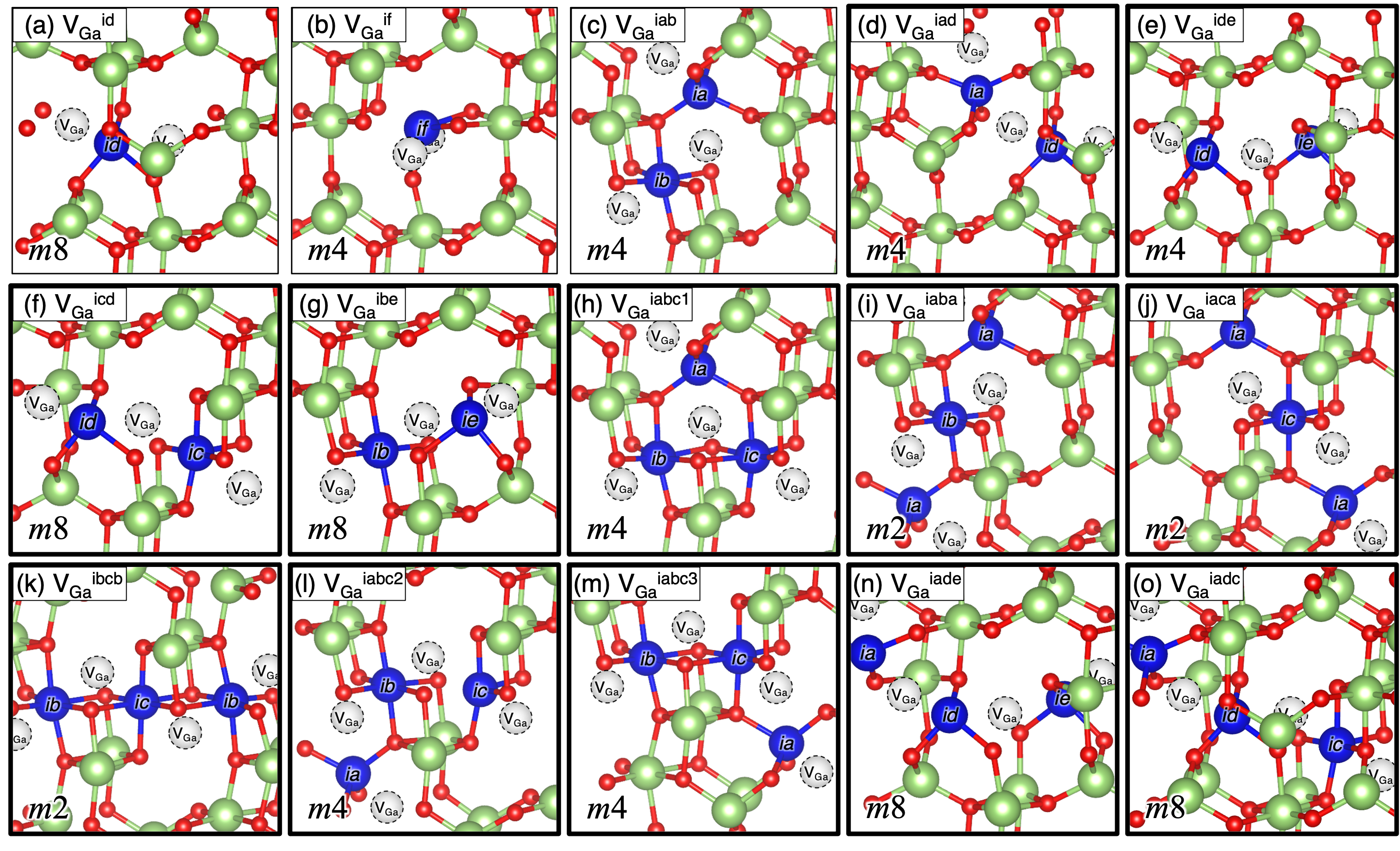}
\caption{
The relaxed $\beta$-Ga$_2$O$_3$ structures with V$_{\text{Ga}}^{3-}$, which are not included in the main text, are also categorized in the same manner as (a,b) double-split, (c-g) triple-split, and (h-o) quadruple-split vacancies. Ga interstitials are represented by blue balls, while Ga vacancies are depicted by gray balls with dashed circles. The Greek letter in the lower left corner indicates the multiplicity of V$_{\text{Ga}}^{3-}$ configurations within the unit cell.
}
\end{figure*}

\newpage

\begin{figure*}[!hbtp] 
\centering
\includegraphics[width=6.5in]{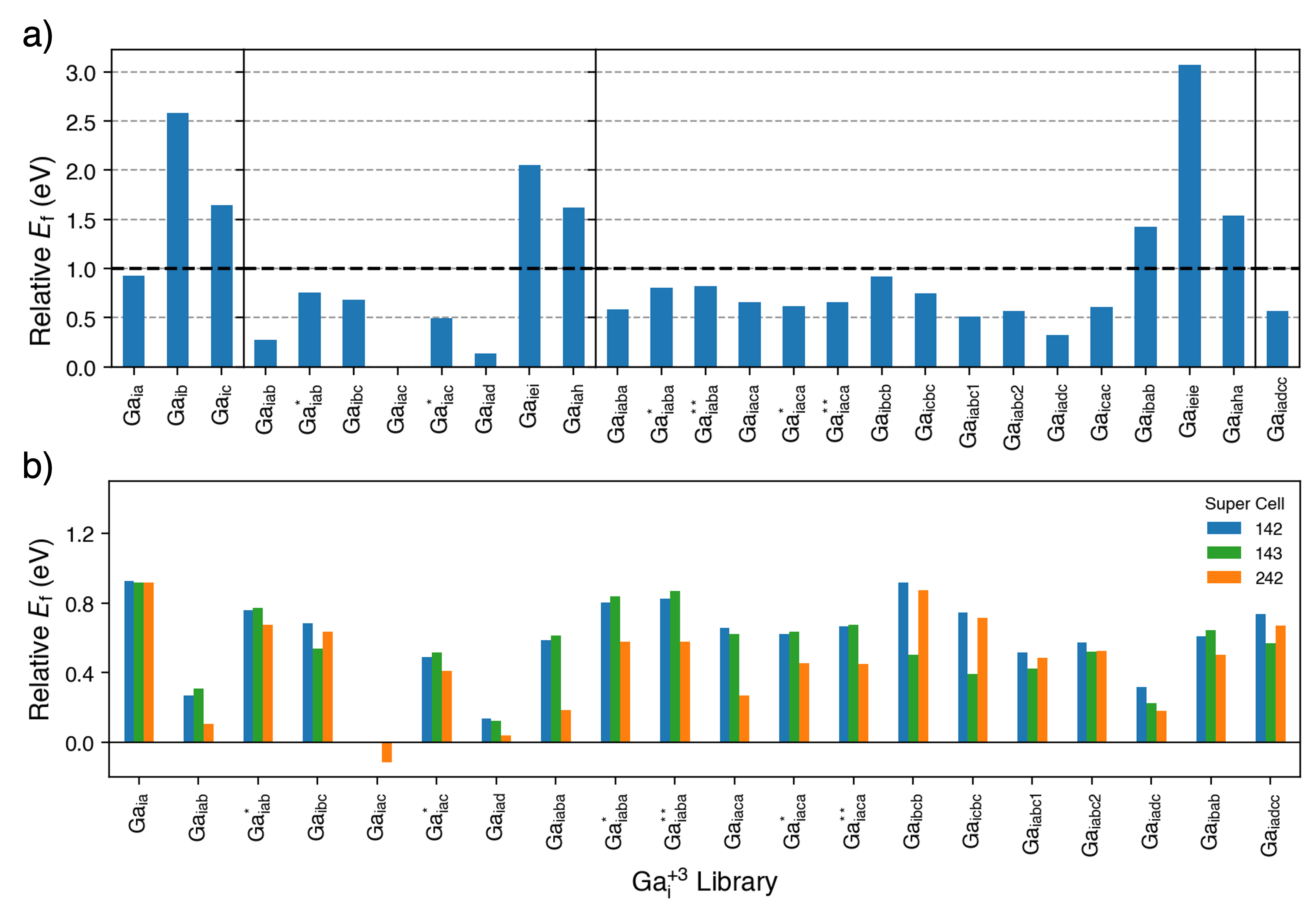}
\caption{
(a) Relative formation energies of 28 different Ga$_{\text{i}}^{3+}$ structures using a 1$\times$4$\times$2 supercell. 
The energies are referenced by the lowest formation energy of Ga$_{\text{iac}}$. 
The energy cut-off to determine inclusion in the hopping network is set at 1 eV.
(b) Relative formation energies of 20 different Ga interstitials down-selected from the full library above to construct the interstitial hopping network under three different supercell sizes (1$\times$4$\times$2, 1$\times$4$\times$3, and 2$\times$4$\times$2), referencing the formation energy of Ga$_{\text{iac}}$ using the 1$\times$4$\times$2 supercell.
}
\end{figure*}

\newpage

\begin{figure*}[!hbtp] 
\centering
\includegraphics[width=6.5in]{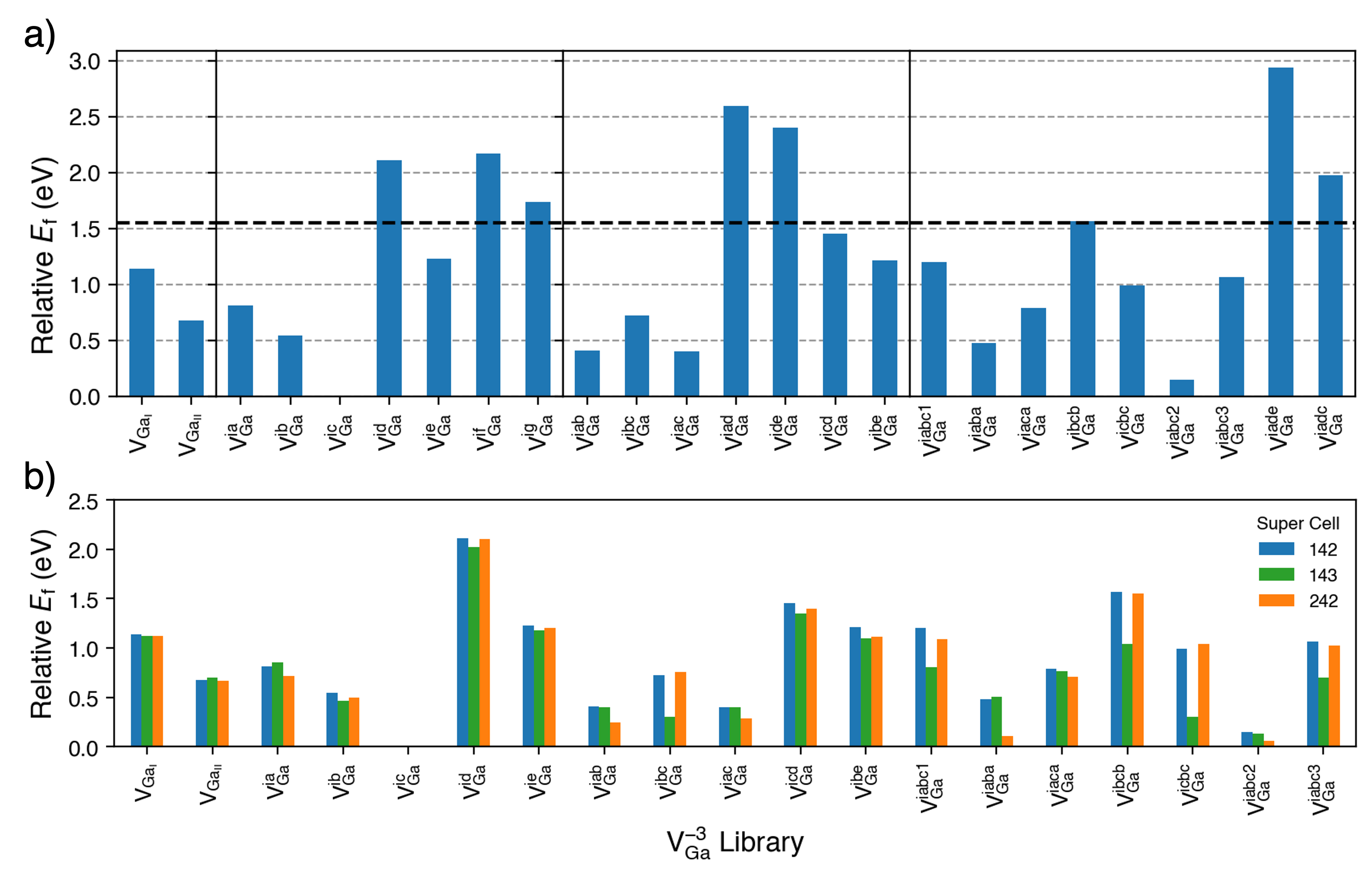}
\caption{
(a) Relative formation energies of 25 different V$_{\text{Ga}}^{3-}$ structures using a 1$\times$4$\times$2 supercell. 
The energies are referenced by the lowest formation energy of V$^{\text{ic}}_{\text{Ga}}$. 
The energy cut-off for determining inclusion in the hopping network is set at 1.55 eV same as the relative formation energy of Q$_4$, V$^{\text{ibcb}}_{\text{Ga}}$.
(b) Relative formation energies of 19 different Ga vacancies down-selected from the full vacancy library above to construct the vacancy hopping network under three different supercell sizes (1$\times$4$\times$2, 1$\times$4$\times$3, and 2$\times$4$\times$2), referencing the formation energy of V$^{\text{ic}}_{\text{Ga}}$ using the 1$\times$4$\times$2 supercell.
}
\end{figure*}

\begin{figure*}[!hbtp] 
\centering
\includegraphics[width=6.5in]{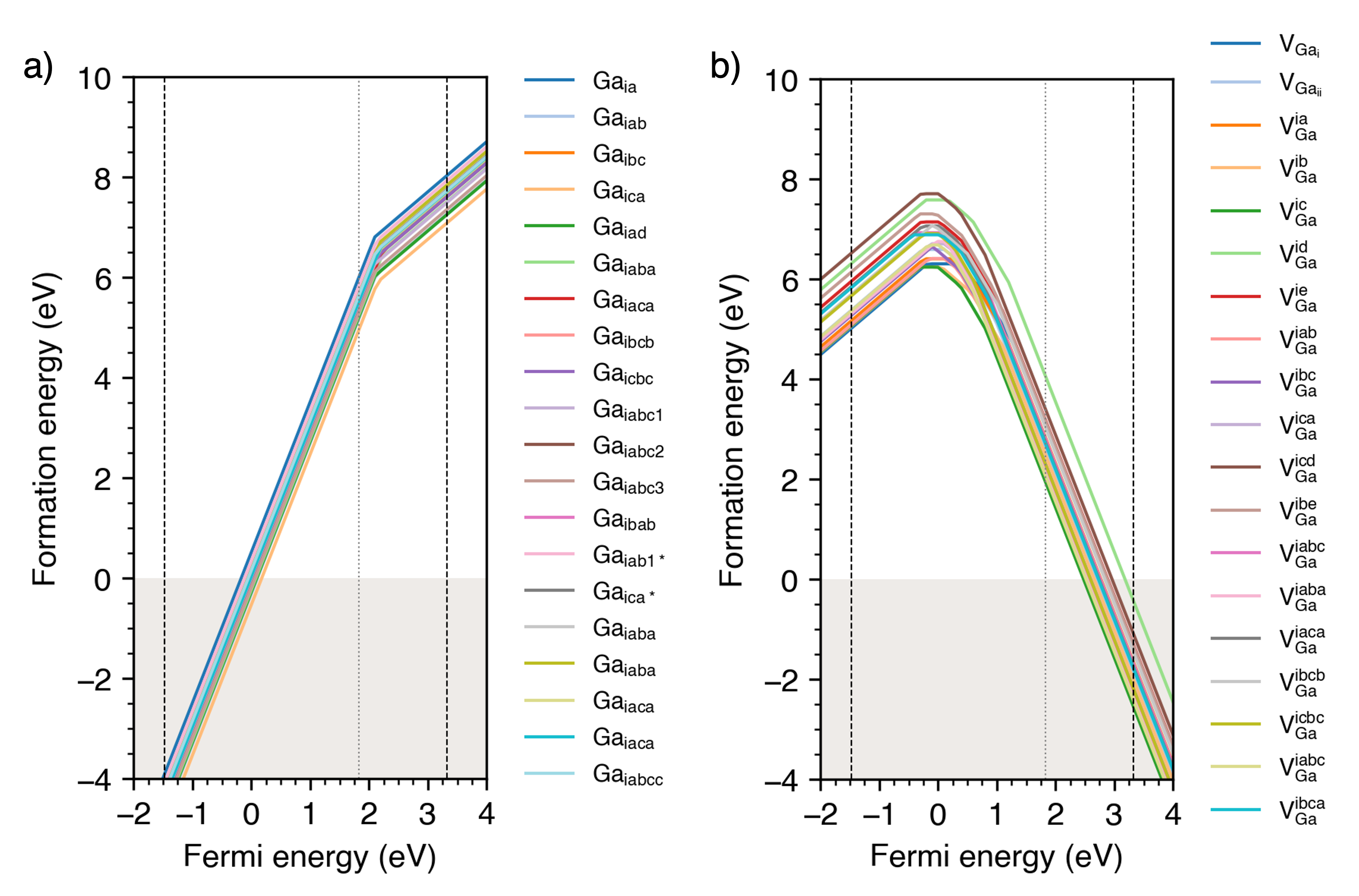}
\caption{
Formation energies of various configurations of (a) Ga interstitials and (b) Ga vacancies as a function of Fermi-energy level using the PBE level of theory under intermediate chemical potential conditions (halfway between the Ga-rich and O-rich limits).
The dotted line represents the Fermi energy at 1.5 eV below CBM.
The dashed lines on the left and right-hand sides of the plot represent the HSE calculated valence band maximum (VBM) and conduction band minimum (CBM) levels, respectively, predicted by band alignments using the electrostatic potentials between PBE and HSE band structures.
}
\end{figure*}

\newpage

\begin{figure*}[!hbtp] 
\centering
\includegraphics[width=6.5in]{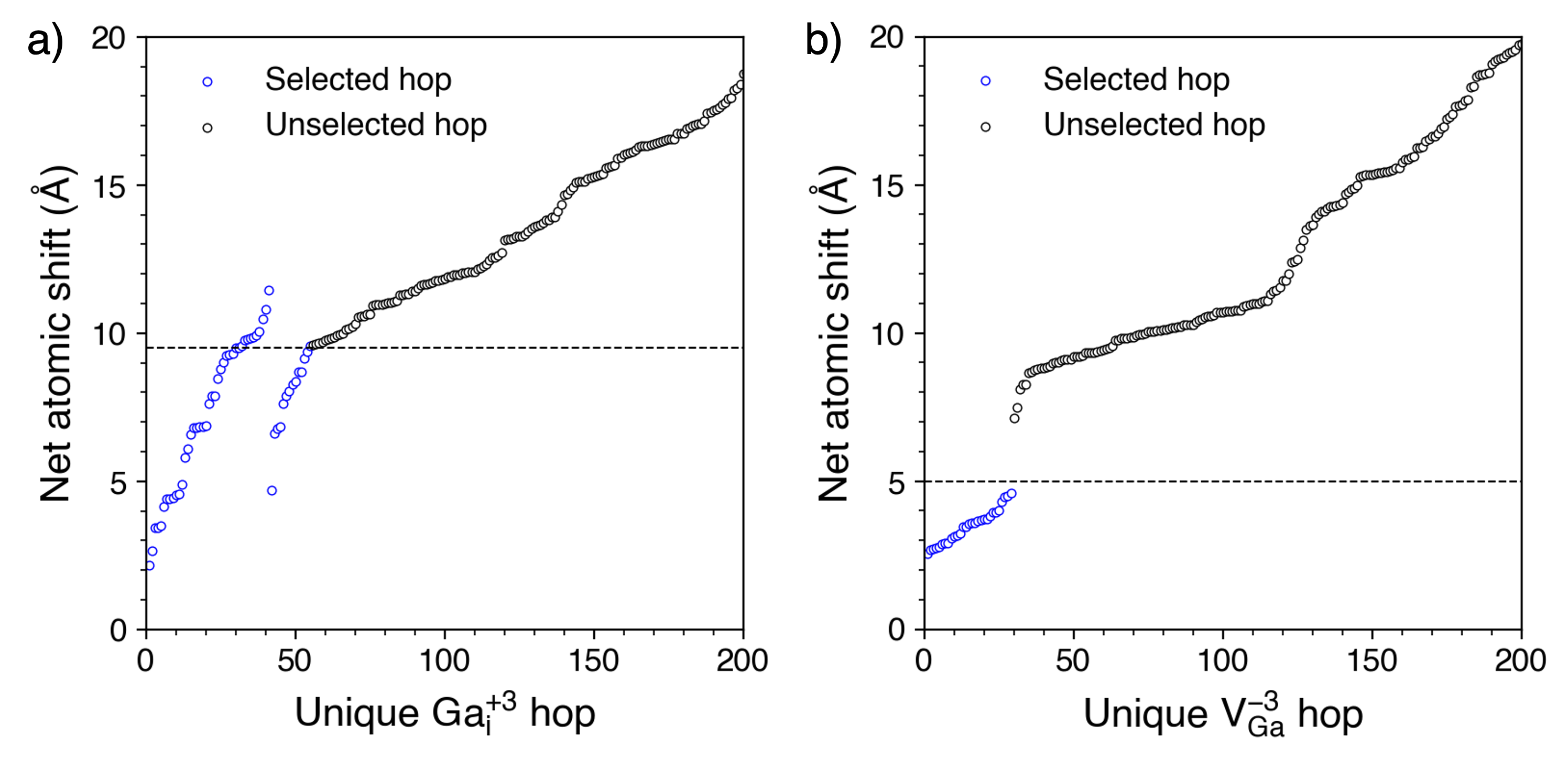}
\caption{
Total atomic displacements of the individual (a) Ga${\text{i}}^{3+}$ hops and (b) V${_{\text{Ga}}^{3-}}$ hops, identified from their respective three-dimensional diffusion networks. The total atomic displacements are calculated by summing the linear trajectories of all atoms within the supercell, from the initial to the final structures of each unique hop.
To reduce the number of hops for further first-principle characterizations, cutoff parameters of 9.7 Å and 5 Å are applied for Ga${_{\text{i}}}^{3+}$ hops and V${_{\text{Ga}}^{3-}}$ hops, respectively, based on their total atomic displacements. Additionally, the number of long-distance atomic trajectories exceeding 1.5 Å is used as an additional cutoff parameter for Ga$_{\text{i}}^{3+}$ hops.
The migration energy barriers of the selected hops are calculated using the CI-NEB methods. Any additional low-energy configurations encountered along the migration path are incorporated into the respective defect library for further analysis, allowing us to isolate and identify each principal hop between the lowest-energy configurations.
}
\end{figure*}

\renewcommand\thetable{S\arabic{table}} 
\newcounter{table1} 
\begin{table}[ht]
\centering
\small
\renewcommand{\arraystretch}{1.2}
\begin{tabular}{cccccc}
\hline
\hline
\multicolumn{1}{c}{Principle hop (PH)} & \multicolumn{1}{c}{Hopping}    & \multicolumn{1}{c}{$\rightharpoonup$ (eV)} & \multicolumn{1}{c}{$\leftharpoondown$ (eV)} & \multicolumn{1}{c}{Supercell} & \multicolumn{1}{c}{Images} \\ \hline
1  & S$_1$. Ga$_{\text{ia}}$  $\rightleftharpoons$   S$_1$. Ga$_{\text{ia}}$  & 0.46  & 0.46  & 142    & 8     \\
2  & S$_1$. Ga$_{\text{ia}}$  $\rightleftharpoons$ D$_3$. Ga$_{\text{iac}}$   & 0.14  & 1.17  & 242    & 4     \\
3  & S$_1$. Ga$_{\text{ia}}$  $\rightleftharpoons$ D$_4$. Ga$_{\text{iad}}$  & 0.12  & 0.92  & 142    & 8     \\
4  & S$_1$. Ga$_{\text{ia}}$  $\rightleftharpoons$ D$_4$. Ga$_{\text{iad}}$  & 0.38  & 1.17  & 142    & 8     \\
5  & S$_1$. Ga$_{\text{ia}}$  $\rightleftharpoons$ T$_8$. Ga$_{\text{ibab}}$  & 0.34  & 0.66  & 142    & 8     \\
6  & S$_1$. Ga$_{\text{ia}}$  $\rightleftharpoons$ D$_{1^{*}}$. Ga$_{\text{iab}^*}$     & 0.15  & 0.32  & 142    & 8     \\
7  & S$_1$. Ga$_{\text{ia}}$  $\rightleftharpoons$ D$_{3^*}$. Ga$_{\text{iac}^*}$ & 0.15  & 0.59  & 142    & 8     \\
8  & D$_1$. Ga$_{\text{iab}}$  $\rightleftharpoons$ D$_3$. Ga$_{\text{iac}}$     & 0.28  & 0.50  & 242    & 4     \\
9  & D$_1$. Ga$_{\text{iab}}$  $\rightleftharpoons$ D$_4$. Ga$_{\text{iad}}$    & 0.14  & 0.21  & 242    & 4     \\
10 & D$_1$. Ga$_{\text{iab}}$  $\rightleftharpoons$ T$_1$. Ga$_{\text{iaba}}$    & 0.37  & 0.29  & 242    & 4     \\
11 & D$_1$. Ga$_{\text{iab}}$  $\rightleftharpoons$ T$_5$. Ga$_{\text{iabc1}}$   & 0.54  & 0.28  & 243    & 4     \\
12 & D$_1$. Ga$_{\text{iab}}$  $\rightleftharpoons$ D$_{1^{*}}$. Ga$_{\text{iab}^*}$       & 0.69  & 0.12  & 242    & 4     \\
13 & D$_2$. Ga$_{\text{ibc}}$  $\rightleftharpoons$ T$_3$. Ga$_{\text{ibcb}}$     & 0.37  & 0.41  & 143    & 8     \\
14 & D$_2$. Ga$_{\text{ibc}}$  $\rightleftharpoons$ T$_4$. Ga$_{\text{icbc}}$     & 0.50  & 0.65  & 143    & 8     \\
15 & D$_2$. Ga$_{\text{ibc}}$  $\rightleftharpoons$ T$_5$. Ga$_{\text{iabc1}}$    & 0.28  & 0.39  & 143    & 8     \\
16 & D$_2$. Ga$_{\text{ibc}}$  $\rightleftharpoons$ T$_6$. Ga$_{\text{iabc2}}$    & 0.30  & 0.31  & 143    & 8     \\
17 & D$_2$. Ga$_{\text{ibc}}$  $\rightleftharpoons$ T$_7$. Ga$_{\text{iadc}}$    & 0.88  & 1.19  & 143    & 8     \\
18 & D$_3$. Ga$_{\text{iac}}$  $\rightleftharpoons$ D$_4$. Ga$_{\text{iad}}$     & 0.17  & 0.02  & 242    & 4     \\
19 & D$_3$. Ga$_{\text{iac}}$  $\rightleftharpoons$ T$_2$. Ga$_{\text{iaca}}$     & 0.64  & 0.26  & 242    & 4     \\
20 & D$_3$. Ga$_{\text{iac}}$  $\rightleftharpoons$ T$_6$. Ga$_{\text{iabc2}}$    & 0.67  & 0.03  & 242    & 4     \\
21 & D$_3$. Ga$_{\text{iac}}$  $\rightleftharpoons$ D$_{3^*}$. Ga$_{\text{iac}^*}$       & 0.69  & 0.17  & 242    & 4     \\
22 & D$_4$. Ga$_{\text{iad}}$  $\rightleftharpoons$ T$_7$. Ga$_{\text{iadc}}$   & 0.20  & 0.10  & 143    & 8     \\
23 & D$_4$. Ga$_{\text{iad}}$  $\rightleftharpoons$ Q$_1$. Ga$_{\text{iadcc}}$   & 0.51  & 0.07  & 143    & 8     \\
24 & T$_1$. Ga$_{\text{iaba}}$  $\rightleftharpoons$ T$_{1^{*}}$. Ga$_{\text{iaba}^{*}}$     & 0.49  & 0.10  & 242    & 4     \\
25 & T$_1$. Ga$_{\text{iaba}}$  $\rightleftharpoons$ T$_{1^{**}}$. Ga$_{\text{iaba}^{**}}$   & 0.53  & 0.13  & 242    & 4     \\
26 & T$_2$. Ga$_{\text{iaca}}$  $\rightleftharpoons$ T$_{2^{*}}$. Ga$_{\text{iaca}^{*}}$     & 0.42  & 0.24  & 242    & 4     \\
27 & T$_2$. Ga$_{\text{iaca}}$  $\rightleftharpoons$ T$_{2^{**}}$. Ga$_{\text{iaca}^{**}}$   & 0.45  & 0.27  & 242    & 4     \\
28 & T$_7$. Ga$_{\text{iadc}}$  $\rightleftharpoons$ Q$_1$. Ga$_{\text{iadcc}}$  & 0.36  & 0.01  & 143    & 8     \\
29 & T$_8$. Ga$_{\text{ibab}}$  $\rightleftharpoons$ D$_{3^*}$. Ga$_{\text{iac}^*}$      & 0.11  & 0.22  & 142    & 8     \\
30 & T$_8$. Ga$_{\text{ibab}}$  $\rightleftharpoons$ Q$_1$. Ga$_{\text{iadcc}}$   & 0.65  & 0.72  & 143    & 8     \\
31 & D$_{1^{*}}$. Ga$_{\text{iab}^*}$  $\rightleftharpoons$ T$_{1^{*}}$. Ga$_{\text{iaba}^{*}}$ & 0.21  & 0.30  & 242    & 4     \\
32 & D$_{3^*}$. Ga$_{\text{iac}^*}$  $\rightleftharpoons$ T$_{2^{*}}$. Ga$_{\text{iaca}^{*}}$    & 0.25  & 0.20  & 242    & 4     \\ \hline
\hline
\end{tabular}
\caption{
A list of principle hops (PHs) between different Ga$_{\text{i}}^{3+}$ configurations, which cannot be further decomposed into multiple shorter elementary hops, and their corresponding hopping barriers in forward ($\rightharpoonup$) and backward ($\leftharpoondown$) directions, respectively. 
The migration hopping barriers are calculated by the CI-NEB using the PBE functional.
The supercell size for CI-NEB calculations was determined to be the minimum size that allows for negligible size effects on formation energies of both initial and final Ga$_{\text{i}}^{3+}$ configurations.
}

\end{table}

\newpage

\begin{table}[!hbtp]
\centering
\small
\renewcommand{\arraystretch}{1.2}
\begin{tabular}{cccccc}
\hline
\hline
\multicolumn{1}{c}{PH} & \multicolumn{1}{c}{Hopping}    & \multicolumn{1}{c}{$\rightharpoonup$ (eV)} & \multicolumn{1}{c}{$\leftharpoondown$ (eV)} & \multicolumn{1}{c}{Supercell} & \multicolumn{1}{c}{Images} \\ \hline
1  & S$_1$. V$_{\text{Ga}_{\text{i}}}$ $\rightleftharpoons$ (D$_6$. V$_{\text{Ga}}^{\text{if}}$) $\rightleftharpoons$ S$_1$. V$_{\text{Ga}_{\text{i}}}$     & 1.07  & 1.07  & 142    & 8     \\
2  & S$_1$. V$_{\text{Ga}_{\text{i}}}$   $\rightleftharpoons$ D$_1$. V$_{\text{Ga}}^{\text{ia}}$  & 0.40  & 0.64  & 142    & 8     \\
3  & S$_1$. V$_{\text{Ga}_{\text{i}}}$   $\rightleftharpoons$ D$_2$. V$_{\text{Ga}}^{\text{ib}}$  & 0.48  & 1.08  & 142    & 8     \\
4  & S$_1$. V$_{\text{Ga}_{\text{i}}}$   $\rightleftharpoons$ D$_3$. V$_{\text{Ga}}^{\text{ic}}$  & 0.31  & 1.45  & 142    & 8     \\
5  & S$_1$. V$_{\text{Ga}_{\text{i}}}$   $\rightleftharpoons$ D$_4$. V$_{\text{Ga}}^{\text{id}}$  & 1.08  & 0.10  & 142    & 8     \\
6  & S$_1$. V$_{\text{Ga}_{\text{i}}}$   $\rightleftharpoons$ D$_5$. V$_{\text{Ga}}^{\text{ie}}$  & 0.32  & 0.22  & 142    & 8     \\
7  & S$_2$. V$_{\text{Ga}_{\text{ii}}}$ $\rightleftharpoons$ (D$_7$. V$_{\text{Ga}}^{\text{ig}}$)  $\rightleftharpoons$ S$_2$. V$_{\text{Ga}_{\text{ii}}}$     & 1.16  & 1.16  & 142    & 8     \\
8  & S$_2$. V$_{\text{Ga}_{\text{ii}}}$   $\rightleftharpoons$ D$_1$. V$_{\text{Ga}}^{\text{ia}}$     & 0.68  & 0.48  & 142    & 8     \\
9  & S$_2$. V$_{\text{Ga}_{\text{ii}}}$   $\rightleftharpoons$ D$_4$. V$_{\text{Ga}}^{\text{id}}$     & 1.88  & 0.44  & 142    & 8     \\
10 & S$_2$. V$_{\text{Ga}_{\text{ii}}}$   $\rightleftharpoons$ D$_5$. V$_{\text{Ga}}^{\text{ie}}$     & 1.13  & 0.57  & 142    & 8     \\
11 & D$_1$. V$_{\text{Ga}}^{\text{ia}}$   $\rightleftharpoons$ T$_1$. V$_{\text{Ga}}^{\text{iab}}$    & 0.42  & 0.90  & 242    & 4     \\
12 & D$_1$. V$_{\text{Ga}}^{\text{ia}}$   $\rightleftharpoons$ T$_3$. V$_{\text{Ga}}^{\text{iac}}$    & 0.51  & 0.93  & 242    & 4     \\
13 & D$_2$. V$_{\text{Ga}}^{\text{ib}}$   $\rightleftharpoons$ T$_1$. V$_{\text{Ga}}^{\text{iab}}$    & 0.40  & 0.66  & 242    & 4     \\
14 & D$_2$. V$_{\text{Ga}}^{\text{ib}}$   $\rightleftharpoons$ T$_2$. V$_{\text{Ga}}^{\text{ibc}}$    & 0.41  & 0.57  & 143    & 8     \\
15 & D$_2$. V$_{\text{Ga}}^{\text{ib}}$   $\rightleftharpoons$ T$_7$. V$_{\text{Ga}}^{\text{ibe}}$    & 0.68  & 0.00  & 142    & 8     \\
16 & D$_3$. V$_{\text{Ga}}^{\text{ic}}$   $\rightleftharpoons$ T$_2$. V$_{\text{Ga}}^{\text{ibc}}$    & 0.86  & 0.56  & 143    & 8     \\
17 & D$_3$. V$_{\text{Ga}}^{\text{ic}}$   $\rightleftharpoons$ T$_3$. V$_{\text{Ga}}^{\text{iac}}$    & 0.63  & 0.34  & 242    & 4     \\
18 & D$_3$. V$_{\text{Ga}}^{\text{ic}}$   $\rightleftharpoons$ T$_6$. V$_{\text{Ga}}^{\text{icd}}$    & 1.48  & 0.02  & 142    & 8     \\
19 & D$_4$. V$_{\text{Ga}}^{\text{id}}$   $\rightleftharpoons$ T$_6$. V$_{\text{Ga}}^{\text{icd}}$    & 0.28  & 0.94  & 142    & 8     \\
20 & D$_5$. V$_{\text{Ga}}^{\text{ie}}$   $\rightleftharpoons$ T$_7$. V$_{\text{Ga}}^{\text{ibe}}$    & 0.67  & 0.69  & 142    & 8     \\
21 & T$_1$. V$_{\text{Ga}}^{\text{iab}}$   $\rightleftharpoons$ Q$_1$. V$_{\text{Ga}}^{\text{iabc1}}$ & 0.86  & 0.39  & 243    & 4     \\
22 & T$_1$. V$_{\text{Ga}}^{\text{iab}}$   $\rightleftharpoons$ Q$_2$. V$_{\text{Ga}}^{\text{iaba}}$  & 0.47  & 0.61  & 242    & 4     \\
23 & T$_1$. V$_{\text{Ga}}^{\text{iab}}$   $\rightleftharpoons$ Q$_6$. V$_{\text{Ga}}^{\text{iabc2}}$ & 0.45  & 0.63  & 242    & 4     \\
24 & T$_2$. V$_{\text{Ga}}^{\text{ibc}}$   $\rightleftharpoons$ Q$_1$. V$_{\text{Ga}}^{\text{iabc1}}$ & 0.79  & 0.28  & 143    & 8     \\
25 & T$_2$. V$_{\text{Ga}}^{\text{ibc}}$   $\rightleftharpoons$ Q$_4$. V$_{\text{Ga}}^{\text{ibcb}}$  & 1.05  & 0.31  & 143    & 4     \\
26 & T$_2$. V$_{\text{Ga}}^{\text{ibc}}$   $\rightleftharpoons$ Q$_5$. V$_{\text{Ga}}^{\text{icbc}}$  & 0.44  & 0.43  & 143    & 4     \\
27 & T$_2$. V$_{\text{Ga}}^{\text{ibc}}$   $\rightleftharpoons$ Q$_6$. V$_{\text{Ga}}^{\text{iabc2}}$ & 0.45  & 0.72  & 243    & 4     \\
28 & T$_2$. V$_{\text{Ga}}^{\text{ibc}}$   $\rightleftharpoons$ Q$_7$. V$_{\text{Ga}}^{\text{iabc3}}$ & 0.63  & 0.22  & 143    & 4     \\
29 & T$_3$. V$_{\text{Ga}}^{\text{iac}}$   $\rightleftharpoons$ Q$_1$. V$_{\text{Ga}}^{\text{iabc1}}$ & 0.83  & 0.46  & 243    & 4     \\
30 & T$_3$. V$_{\text{Ga}}^{\text{iac}}$   $\rightleftharpoons$ Q$_3$. V$_{\text{Ga}}^{\text{iaca}}$  & 0.68  & 0.26  & 242    & 4     \\
31 & T$_3$. V$_{\text{Ga}}^{\text{iac}}$   $\rightleftharpoons$ Q$_7$. V$_{\text{Ga}}^{\text{iabc3}}$ & 0.91  & 0.53  & 243    & 4     \\ \hline
\hline
\end{tabular}
\caption{
A list of PHs between different V$_{\text{Ga}}^{3-}$ configurations, which cannot be further decomposed into multiple shorter elementary hops, and corresponding hopping barriers in forward ($\rightharpoonup$) and backward ($\leftharpoondown$) directions, respectively. 
The migration energy barriers are calculated by the CI-NEB using the PBE functional.
The supercell size for CI-NEB calculations was determined to be the minimum size that allows for negligible size effects on formation energies of both initial and final V$_{\text{Ga}}^{3-}$ configurations.
}

\end{table}

\newpage

\begin{table}[!hbtp]
\renewcommand{\arraystretch}{1.2}
\centering
\small
\begin{tabular}{cccccl}
\hline
\hline
\multicolumn{1}{c}{Hopping}  & \multicolumn{1}{c}{$D_{\text{Mid.'s}}$ (\AA)} & \multicolumn{1}{c}{$\Sigma \Delta l$ (\AA)} & Large $\Delta l$  & \multicolumn{1}{c}{PH} & \multicolumn{1}{c}{Hopping combination}     \\ \hline
D$_4$. Ga$_{\text{iad}}$ $\rightleftharpoons$ T$_7$. Ga$_{\text{iadc}}$ & 1.58 & 2.18  & 0  & 22  &        \\
T$_8$. Ga$_{\text{ibab}}$   $\rightleftharpoons$ D$_{3^*}$. Ga$_{\text{iac}^*}$    & 1.16 & 2.64  & 0  & 29   &        \\
D$_1$. Ga$_{\text{iab}}$   $\rightleftharpoons$ T$_5$. Ga$_{\text{iabc1}}$ & 1.48 & 3.43  & 0  & 11   &        \\
D$_1$. Ga$_{\text{iab}}$   $\rightleftharpoons$ D$_{1^{*}}$. Ga$_{\text{iab}^*}$     & 0.47 & 3.43  & 0  & 12   &        \\
D$_3$. Ga$_{\text{iac}}$   $\rightleftharpoons$ T$_6$. Ga$_{\text{iabc2}}$  & 1.55 & 3.50  & 0  & 20   &        \\
D$_2$. Ga$_{\text{ibc}}$   $\rightleftharpoons$ T$_5$. Ga$_{\text{iabc1}}$  & 1.58 & 4.17  & 0  & 15   &        \\
D$_1$. Ga$_{\text{iab}}$   $\rightleftharpoons$ T$_1$. Ga$_{\text{iaba}}$  & 1.79 & 4.40  & 0  & 10  &        \\
D$_2$. Ga$_{\text{ibc}}$   $\rightleftharpoons$ T$_3$. Ga$_{\text{ibcb}}$   & 1.55 & 4.41  & 0  & 13  &        \\
D$_2$. Ga$_{\text{ibc}}$   $\rightleftharpoons$ T$_6$. Ga$_{\text{iabc2}}$  & 1.60 & 4.44  & 0  & 16  &        \\
D$_3$. Ga$_{\text{iac}}$   $\rightleftharpoons$ D$_4$. Ga$_{\text{iad}}$   & 3.03 & 4.54  & 0  & 18  &        \\
D$_3$. Ga$_{\text{iac}}$   $\rightleftharpoons$ D$_{3^*}$. Ga$_{\text{iac}^*}$     & 0.54 & 4.58  & 0  & 21  &        \\
D$_3$. Ga$_{\text{iac}}$   $\rightleftharpoons$ T$_2$. Ga$_{\text{iaca}}$   & 1.84 & 4.88  & 0  & 19  &        \\
S$_1$. Ga$_{\text{ia}}$   $\rightleftharpoons$ D$_{1^{*}}$. Ga$_{\text{iab}^*}$       & 1.86 & 5.80  & 0  &  6  &        \\
D$_1$. Ga$_{\text{iab}}$   $\rightleftharpoons$ D$_4$. Ga$_{\text{iad}}$  & 2.74 & 6.08  & 0  &  9  &        \\
S$_1$. Ga$_{\text{ia}}$   $\rightleftharpoons$ D$_{3^*}$. Ga$_{\text{iac}^*}$      & 1.85 & 6.59  & 0  &  7  &        \\
T$_1$. Ga$_{\text{iaba}}$   $\rightleftharpoons$ D$_{1^{*}}$. Ga$_{\text{iab}^*}$     & 1.76 & 6.81  & 0  & $\times$   & T$_1$ $\rightleftharpoons$ D$_1$   $\rightleftharpoons$ D$_{1^{*}}$       \\
T$_5$. Ga$_{\text{iabc1}}$   $\rightleftharpoons$ D$_{1^{*}}$. Ga$_{\text{iab}^*}$    & 1.54 & 6.82  & 0  & $\times$   & T$_5$ $\rightleftharpoons$ D$_1$   $\rightleftharpoons$ D$_{1^{*}}$       \\
T$_6$. Ga$_{\text{iabc2}}$   $\rightleftharpoons$ D$_{3^*}$. Ga$_{\text{iac}^*}$       & 1.48 & 6.85  & 0  & $\times$   & T$_6$ $\rightleftharpoons$ D$_{1^{*}}$   $\rightleftharpoons$ D$_{3^*}$       \\
S$_1$. Ga$_{\text{ia}}$   $\rightleftharpoons$ D$_4$. Ga$_{\text{iad}}$    & 2.53 & 6.85  & 0  & 3  &        \\
T$_2$. Ga$_{\text{iaca}}$   $\rightleftharpoons$ D$_{3^*}$. Ga$_{\text{iac}^*}$    & 1.63 & 6.89  & 0  & $\times$   & T$_2$ $\rightleftharpoons$ D$_3$   $\rightleftharpoons$ D$_{3^*}$         \\
S$_1$. Ga$_{\text{ia}}$   $\rightleftharpoons$ D$_1$. Ga$_{\text{iab}}$    & 2.54 & 7.63  & 0  & $\times$   & S$_1$ $\rightleftharpoons$ D$_4$   $\rightleftharpoons$ D$_1$    \\
S$_1$. Ga$_{\text{ia}}$   $\rightleftharpoons$ T$_8$. Ga$_{\text{ibab}}$    & 1.79 & 7.88  & 0  &  5  &        \\
S$_1$. Ga$_{\text{ia}}$   $\rightleftharpoons$ D$_3$. Ga$_{\text{iac}}$     & 2.03 & 7.90  & 0  &  2  &        \\
D$_2$. Ga$_{\text{ibc}}$   $\rightleftharpoons$ D$_{3^*}$. Ga$_{\text{iac}^*}$     & 2.99 & 8.48  & 0  & $\times$   & D$_2$ $\rightleftharpoons$ D$_3$   $\rightleftharpoons$ D$_{3^*}$         \\
D$_{3^*}$. Ga$_{\text{iac}^*}$   $\rightleftharpoons$ D$_{3^*}$. Ga$_{\text{iac}^*}$   & 3.15 & 8.78  & 0  & $\times$   & D$_{3^*}$ $\rightleftharpoons$   (T$_{2^{*}}$,T$_{2^{**}}$) $\rightleftharpoons$ D$_{3^*}$         \\
D$_1$. Ga$_{\text{iab}}$   $\rightleftharpoons$ T$_7$. Ga$_{\text{iadc}}$ & 3.89 & 9.03  & 0  & $\times$   & D$_1$ $\rightleftharpoons$ D$_3$   $\rightleftharpoons$ T$_7$    \\
D$_1$. Ga$_{\text{iab}}$   $\rightleftharpoons$ D$_{1^{*}}$. Ga$_{\text{iab}^*}$     & 3.52 & 9.23  & 0  & $\times$   & D$_1$ $\rightleftharpoons$ T$_1$   $\rightleftharpoons$ D$_1$ $\rightleftharpoons$ D$_{1^{*}}$     \\
D$_{1^{*}}$. Ga$_{\text{iab}^*}$   $\rightleftharpoons$ D$_{1^{*}}$. Ga$_{\text{iab}^*}$ & 3.53 & 9.29  & 0  & $\times$   & D$_{1^{*}}$ $\rightleftharpoons$   (T$_{1^{*}}$,T$_{1^{**}}$) $\rightleftharpoons$ D$_{1^{*}}$     \\
D$_3$. Ga$_{\text{iac}}$   $\rightleftharpoons$ D$_{1^{*}}$. Ga$_{\text{iab}^*}$      & 3.70 & 9.30  & 0  & $\times$   & D$_3$ $\rightleftharpoons$ D$_4$   $\rightleftharpoons$ S$_1$ $\rightleftharpoons$ D$_{1^{*}}$     \\
D$_{1^{*}}$. Ga$_{\text{iab}^*}$   $\rightleftharpoons$ D$_{1^{*}}$. Ga$_{\text{iab}^*}$ & 2.20 & 9.51  & 0  & $\times$   & D$_{1^{*}}$ $\rightleftharpoons$ S$_1$   $\rightleftharpoons$ D$_{1^{*}}$     \\
D$_4$. Ga$_{\text{iad}}$   $\rightleftharpoons$ D$_{3^*}$. Ga$_{\text{iac}^*}$    & 3.34 & 9.52  & 0  & $\times$   & D$_4$ $\rightleftharpoons$ D$_3$   $\rightleftharpoons$ D$_{3^*}$         \\
D$_{1^{*}}$. Ga$_{\text{iab}^*}$   $\rightleftharpoons$ D$_{3^*}$. Ga$_{\text{iac}^*}$    & 2.35 & 9.58  & 0  & $\times$   & D$_{1^{*}}$ $\rightleftharpoons$ S$_1$   $\rightleftharpoons$ D$_3$ $\rightleftharpoons$ D$_{3^*}$     \\
\hline
\hline
\end{tabular}
\caption{
A selected list of unique hops found in three-dimensional Ga$_{\text{i}}^{3+}$ diffusion network shown in Figure S6, sorted by the number of long atomic shifts over 1.5 {\AA} and the total atomic shifts across all atoms within the supercell.
}

\end{table}

\newpage

\begin{table}[!hbtp]
\renewcommand{\arraystretch}{1.2}
\centering
\small
\begin{tabular}{cccccl}
\hline
\hline
\multicolumn{1}{c}{Hopping}  & \multicolumn{1}{c}{$D_{\text{Mid.'s}}$ (\AA)} & \multicolumn{1}{c}{$\Sigma \Delta l$ (\AA)} & Large $\Delta l$  & \multicolumn{1}{c}{PH} & \multicolumn{1}{c}{Hopping combination}     \\ \hline
D$_1$. Ga$_{\text{iab}}$   $\rightleftharpoons$ D$_1$. Ga$_{\text{iab}}$  & 3.57 & 9.77  & 0  & $\times$   & D$_1$ $\rightleftharpoons$ T$_1$   $\rightleftharpoons$ D$_1$    \\
D$_{3^*}$. Ga$_{\text{iac}^*}$   $\rightleftharpoons$ D$_{3^*}$. Ga$_{\text{iac}^*}$   & 3.26 & 9.80  & 0  & $\times$   & D$_{3^*}$ $\rightleftharpoons$   (T$_{2^{*}}$,T$_{2^{**}}$) $\rightleftharpoons$ D$_{3^*}$         \\
D$_{1^{*}}$. Ga$_{\text{iab}^*}$   $\rightleftharpoons$ D$_{1^{*}}$. Ga$_{\text{iab}^*}$ & 3.41 & 9.83  & 0  & $\times$   & D$_{1^{*}}$ $\rightleftharpoons$   (T$_{1^{*}}$,T$_{1^{**}}$) $\rightleftharpoons$ D$_{1^{*}}$     \\
D$_4$. Ga$_{\text{iad}}$   $\rightleftharpoons$ D$_{1^{*}}$. Ga$_{\text{iab}^*}$     & 3.34 & 9.87  & 0  & $\times$   & D$_4$ $\rightleftharpoons$ S$_1$   $\rightleftharpoons$ D$_{1^{*}}$       \\
D$_4$. Ga$_{\text{iad}}$   $\rightleftharpoons$ D$_{1^{*}}$. Ga$_{\text{iab}^*}$     & 2.94 & 9.95  & 0  & $\times$   & D$_4$ $\rightleftharpoons$ D$_1$   $\rightleftharpoons$ D$_{1^{*}}$       \\
D$_1$. Ga$_{\text{iab}}$   $\rightleftharpoons$ D$_2$. Ga$_{\text{ibc}}$   & 3.04 & 10.06     & 0  & $\times$   & D$_1$ $\rightleftharpoons$ T$_5$   $\rightleftharpoons$ D$_2$    \\
D$_4$. Ga$_{\text{iad}}$   $\rightleftharpoons$ T$_5$. Ga$_{\text{iabc1}}$ & 3.32 & 10.50     & 0  & $\times$   & D$_4$ $\rightleftharpoons$ D$_1$   $\rightleftharpoons$ T$_5$    \\
D$_{3^*}$. Ga$_{\text{iac}^*}$   $\rightleftharpoons$ D$_{3^*}$. Ga$_{\text{iac}^*}$   & 2.24 & 10.82     & 0  & $\times$   & D$_{3^*}$ $\rightleftharpoons$ T$_8$   $\rightleftharpoons$ D$_{3^*}$         \\
S$_1$. Ga$_{\text{ia}}$   $\rightleftharpoons$ T$_5$. Ga$_{\text{iabc1}}$   & 3.95 & 11.45     & 0  & $\times$   & S$_1$ $\rightleftharpoons$ D$_4$   $\rightleftharpoons$ D$_1$ $\rightleftharpoons$ T$_5$       \\
D$_2$. Ga$_{\text{ibc}}$   $\rightleftharpoons$ T$_4$. Ga$_{\text{icbc}}$   & 1.46 & 4.70  & 1  & 14   &        \\
D$_1$. Ga$_{\text{iab}}$   $\rightleftharpoons$ D$_3$. Ga$_{\text{iac}}$   & 0.84 & 6.63  & 1  & 8   &        \\
D$_{1^{*}}$. Ga$_{\text{iab}^*}$   $\rightleftharpoons$ D$_{1^{*}}$. Ga$_{\text{iab}^*}$ & 0.89 & 6.78  & 1  & $\times$   & D$_{1^{*}}$ $\rightleftharpoons$ D$_1$   $\rightleftharpoons$ D$_{1^{*}}$     \\
D$_3$. Ga$_{\text{iac}}$   $\rightleftharpoons$ T$_8$. Ga$_{\text{ibab}}$   & 1.67 & 6.84  & 1  & $\times$   & D$_3$ $\rightleftharpoons$ D$_{3^*}$   $\rightleftharpoons$ T$_8$         \\
S$_1$. Ga$_{\text{ia}}$   $\rightleftharpoons$ S$_1$. Ga$_{\text{ia}}$  & 1.96 & 7.62  & 1  & 1  &        \\
D$_{3^*}$. Ga$_{\text{iac}^*}$   $\rightleftharpoons$ D$_{3^*}$. Ga$_{\text{iac}^*}$   & 0.85 & 7.89  & 1  & $\times$   & D$_{3^*}$ $\rightleftharpoons$ T$_8$   $\rightleftharpoons$ D$_{3^*}$         \\
S$_1$. Ga$_{\text{ia}}$   $\rightleftharpoons$ D$_4$. Ga$_{\text{iad}}$    & 2.57 & 8.05  & 1  & 4  &        \\
S$_1$. Ga$_{\text{ia}}$   $\rightleftharpoons$ D$_3$. Ga$_{\text{iac}}$     & 1.92 & 8.28  & 1  & $\times$  & S$_1$ $\rightleftharpoons$ D$_{3^{*}}$ $\rightleftharpoons$ D$_3$ \\
S$_1$. Ga$_{\text{ia}}$   $\rightleftharpoons$ D$_1$. Ga$_{\text{iab}}$    & 2.08 & 8.36  & 1  & $\times$   & S$_1$ $\rightleftharpoons$ D$_{1^{*}}$   $\rightleftharpoons$ D$_1$       \\
D$_2$. Ga$_{\text{ibc}}$   $\rightleftharpoons$ D$_{3^*}$. Ga$_{\text{iac}^*}$     & 1.09 & 8.69  & 1  & $\times$   & \begin{tabular}[c]{@{}l@{}}D$_2$ $\rightleftharpoons$ T$_6$   $\rightleftharpoons$ D$_3$ $\rightleftharpoons$ \\ T$_2$ $\rightleftharpoons$   D$_3$ $\rightleftharpoons$ D$_{3^*}$\end{tabular} \\
D$_2$. Ga$_{\text{ibc}}$   $\rightleftharpoons$ D$_3$. Ga$_{\text{iac}}$    & 3.14 & 9.16  & 1  & $\times$   & D$_2$ $\rightleftharpoons$ T$_6$   $\rightleftharpoons$ D$_3$    \\
D$_2$. Ga$_{\text{ibc}}$   $\rightleftharpoons$ T$_7$. Ga$_{\text{iadc}}$  & 2.32 & 9.57  & 1  & 17  &        \\
D$_{1^{*}}$. Ga$_{\text{iab}^*}$   $\rightleftharpoons$ D$_{3^*}$. Ga$_{\text{iac}^*}$    & 0.78 & 9.68  & 1  & $\times$   & D$_{1^{*}}$ $\rightleftharpoons$ S$_1$   $\rightleftharpoons$ D$_{3^*}$       \\ 
\hline
\hline
\end{tabular}
\end{table}